\documentclass[a4paper,11pt]{article}
\pdfoutput=1 

\usepackage{jcappub} 

\usepackage[T1]{fontenc} 

\usepackage{braket}
\usepackage{adjustbox}
\usepackage{graphicx,epsfig}
\usepackage{amsmath}
\usepackage {amssymb}
\usepackage {longtable}
\usepackage{multirow}
 \usepackage{mathrsfs}
\usepackage{float}
\usepackage{dcolumn}
\usepackage{bm}
\usepackage{amsfonts}
\usepackage[normalem]{ulem}
\usepackage{subfigure}
\usepackage{color}

\usepackage{relsize}
\usepackage{booktabs}
\usepackage[final]{microtype}

\newcommand{\be}{\begin{equation}}
\newcommand{\ee}{\end{equation}}
\newcommand{\bea}{\begin{eqnarray}}
\newcommand{\eea}{\end{eqnarray}}

\title{\boldmath Primordial black hole production in scalar field inflation within $f(T)$ gravity}

\author[a]{Daniel Villalobos-Silva}
\author[a]{, Yerko V\'asquez}
\author[b]{and Giovanni Otalora}

\affiliation[a]{Departamento de F\'isica, Facultad de Ciencias, Universidad de La Serena, Avenida Cisternas 1200, La Serena, Chile}
\affiliation[b]{Departamento de F\'isica, Facultad de Ciencias, Universidad de Tarapac\'a, Casilla 7-D, Arica, Chile}

\emailAdd{daniel.villalobos@userena.cl}
\emailAdd{yvasquez@userena.cl}
\emailAdd{giovanni.otalora@academicos.uta.cl}

\abstract{We investigate inflation in modified teleparallel gravity within a scalar--tensor framework.
We focus on two viable extensions of the Teleparallel Equivalent of General Relativity: a power--law model and an exponential model, which introduce controlled deviations from standard teleparallel gravity through a correction parameter $\alpha$.
Inflation is driven by a string--inspired fiber inflation potential that naturally realizes a transient ultra slow--roll (USR) phase.
We solve the background equations numerically and compute the evolution of cosmological perturbations within the modified teleparallel framework.
We show that both models generate an amplification of the primordial curvature power spectrum on small scales due to the USR phase, while remaining compatible with cosmic microwave background constraints at large scales.
The modified gravity sector introduces corrections to the slow--roll parameters, tensor spectral index, and tensor--to--scalar ratio through derivatives of the torsion function, leading to potentially observable signatures distinct from canonical inflation.
We further analyze the implications of enhanced scalar perturbations for primordial black hole (PBH) formation and demonstrate that modified teleparallel gravity provides a theoretically consistent and phenomenologically rich framework for producing PBHs during inflation.
}

\begin{document}

\maketitle
\flushbottom

\section{Introduction}\label{Introduction}

The inflationary paradigm provides a compelling explanation for the observed homogeneity, isotropy, and spatial flatness of the Universe, while simultaneously furnishing a mechanism for the generation of primordial perturbations \cite{Guth:1980zm, Starobinsky:1980te, Linde:1981mu, Albrecht:1982wi}. 
In this framework, quantum fluctuations generated during inflation are stretched beyond the Hubble radius and become imprinted as primordial curvature perturbations. After horizon re-entry, these perturbations seed the formation of large-scale structure and source the temperature anisotropies observed in the cosmic microwave background (CMB), which have been measured with high precision by the Planck satellite \cite{Planck:2018jri}.
The observed perturbations are predominantly adiabatic and nearly scale invariant, in close agreement with the predictions of the simplest inflationary scenarios. 
Nevertheless, the fundamental physical mechanism responsible for driving inflation remains unknown, providing strong motivation to explore extensions beyond General Relativity (GR) \cite{Martin:2013tda,Clifton:2011jh,Nojiri:2010wj,Roy:2026vwx,Kumar:2025apf}.

As an alternative geometric formulation of General Relativity, the Teleparallel Equivalent of General Relativity (TEGR), hereafter simply referred to as Teleparallel Gravity (TG), describes the gravitational interaction in terms of torsion rather than curvature \cite{Einstein,TranslationEinstein, Aldrovandi:2013wha, JGPereira2, AndradeGuillenPereira-00, Arcos:2005ec, Bahamonde:2021gfp}. From a gauge-theoretic perspective, TG can be understood as gauge theory of spacetime translations and is closely related to the broader class of gauge theories of gravity, including Poincaré gauge theory, where both translations and Lorentz transformations are gauged \cite{Aldrovandi:2013wha,Pereira:2019woq,Hehl:1976kj,Blagojevic:2002du,Blagojevic:2013xpa}. While Poincaré gauge gravity generally allows both curvature and torsion, TG corresponds to a curvature-free, metric-compatible formulation in which the gravitational degrees of freedom are carried by torsion. The fundamental variable is the tetrad field, which acts as the gauge potential associated with local translations, provides the soldering between the tangent space and the spacetime manifold, and locally determines the spacetime metric. The gravitational dynamics is then formulated in terms of the torsion tensor of the Weitzenb\"ock connection, in contrast to the standard formulation of GR, where gravity is encoded in the curvature of the torsion-free Levi-Civita connection.

Although TG is dynamically equivalent to GR at the level of the field equations \cite{Hayashi:1979qx, Maluf:2013gaa}, it provides a distinct geometric and gauge-theoretic foundation from which modified theories of gravity can be constructed.  In this framework, the gravitational action is constructed from the torsion scalar $T$, which differs from the Ricci scalar $R$ only by a boundary term, ensuring that both formulations lead to identical classical dynamics \cite{Aldrovandi:2013wha}. This equivalence makes TG a natural starting point for extensions beyond GR. In particular, $f(T)$ gravity arises by promoting the torsion scalar in the action to an arbitrary function 
$f(T)$, leading to richer phenomenology while preserving second-order field equations. Such models have been extensively explored in cosmological contexts, especially in relation to both early- and late-time accelerated expansion \cite{Ferraro:2006jd, Ferraro:2008ey,Fiorini:2013hva, Bohmer:2019qfi, Bahamonde:2021gfp}. In addition to $f(T)$ gravity, more elaborate extensions have been proposed by incorporating additional degrees of freedom and nonminimal couplings. A well-studied class is provided by scalar-torsion models, described by functions $f(T,\phi)$, in which a scalar field is nonminimally coupled to torsion. These theories have been extensively investigated in cosmological contexts, including cosmic inflation \cite{Gonzalez-Espinoza:2020azh,Gonzalez-Espinoza:2021qnv,Leyva:2021fuo,Kumar:2025apf} and dark energy \cite{Gonzalez-Espinoza:2020jss,Gonzalez-Espinoza:2021mwr,Rodriguez-Benites:2024pce}, as well as in compact-object configurations, including black-hole solutions \cite{Geng:2011aj,Gonzalez:2014pwa,Gonzalez:2024ifp}.

Beyond scalar–torsion theories, further generalizations arise by following a route analogous to that of curvature-based modified gravity, where higher-order and higher-derivative invariants may be introduced in the gravitational action. In the teleparallel framework, this motivates the construction of torsional counterparts of such curvature corrections.  A notable example is the teleparallel Gauss–Bonnet framework, where the invariant $T_G$ plays the role of the curvature Gauss–Bonnet term, leading to theories of the form $f(T,T_G)$ with dynamics beyond standard $f(T)$ gravity \cite{Kofinas:2014owa}. This construction can be further extended to teleparallel Lovelock theories, where a hierarchy of torsion invariants generalizes the usual Lovelock scalars and provides a systematic framework for higher-order gravitational actions \cite{Gonzalez:2015sha, Gonzalez:2019tky}. In the same spirit, higher-derivative teleparallel models have been proposed by introducing derivative torsion operators, giving rise to Lagrangians of the form $f(T,(\nabla T)^2,\Box T)$ and leading to nontrivial cosmological dynamics \cite{Otalora:2016dxe}. Higher-order teleparallel corrections have also been analyzed in the weak-field regime through the parametrized post-Newtonian formalism, providing complementary constraints on this class of theories \cite{Gonzalez-Espinoza:2021nqd}. Furthermore, teleparallel Horndeski gravity represents the most general scalar–tensor extension with second-order field equations, containing standard Horndeski theory as a limit while allowing for additional torsion-induced couplings \cite{Bahamonde:2019shr}.

In the context of early-Universe cosmology, modified teleparallel gravity introduces additional degrees of freedom that can significantly affect both the background evolution and the dynamics of cosmological perturbations. 
In particular, scalar–torsion extensions described by functions of the form $f(T,\phi)$, supplemented by the canonical kinetic term $X$ and the scalar potential $V(\phi)$, provide a natural framework to explore deviations from canonical single-field inflation. In these models, the scalar field can drive inflation, while torsional modifications alter the effective gravitational dynamics, potentially leading to distinctive predictions for inflationary observables such as the scalar spectral index, the tensor-to-scalar ratio, and the running of the scalar spectral index \cite{Gonzalez-Espinoza:2020azh,Gonzalez-Espinoza:2021qnv,Leyva:2021fuo,Kumar:2025apf}. 
The generation of primordial perturbations in these theories is, however, sensitive to the precise form of the coupling between the scalar field and torsion. In particular, models with a nonminimal coupling function that is linear in $T$ may face difficulties in producing viable primordial density fluctuations due to the effects of local Lorentz-symmetry breaking \cite{Gonzalez-Espinoza:2020azh,Wu:2016dkt,Raatikainen:2019qey}. This obstruction is not generic, however, as more general scalar–torsion theories with nonlinear dependence on $T$, including suitable $f(T,\phi)$ constructions, can generate consistent scalar and tensor perturbation spectra \cite{Gonzalez-Espinoza:2020azh}. Consequently, modified teleparallel inflation remains a viable framework, provided that the torsional couplings are suitably constructed to ensure a consistent perturbation sector and inflationary predictions compatible with current observational constraints \cite{Gonzalez-Espinoza:2020azh,Gonzalez-Espinoza:2021qnv,Leyva:2021fuo,Kumar:2025apf}.

One of the most intriguing consequences of such models, for appropriate choices of the scalar potential, is the possibility of realizing an ultra slow-roll (USR) phase of inflation, during which the inflaton evolves across an extremely flat region of the potential and the standard slow-roll conditions are temporarily violated \cite{Namjoo:2012aa,Martin:2012pe,Dimopoulos:2017ged, Motohashi:2017kbs}. During this phase, the friction-dominated dynamics can lead to a rapid superhorizon growth of curvature perturbations, resulting in a significant amplification of the primordial power spectrum at small scales. This enhancement can, in turn, trigger the formation of primordial black holes (PBHs) upon horizon re-entry, whose abundance is highly sensitive to the amplitude of the curvature perturbations on scales far smaller than those probed by the cosmic microwave background \cite{Hawking:1971ei, Carr:1974nx, Carr:1975qj,Ballesteros:2017fsr,Sasaki:2018dmp}.

Furthermore, PBHs have attracted considerable interest as viable dark matter candidates \cite{Chapline:1975ojl, Carr:2016drx, Green:2020jor}. Their possible contribution to the dark matter abundance has been extensively investigated, with current observational constraints spanning a wide range of masses and relying on multiple astrophysical and cosmological probes, including microlensing surveys, gravitational wave observations, and limits from Hawking radiation \cite{Carr:2020gox, Green:2020jor,Villanueva-Domingo:2021spv}. As a result, the PBH abundance is tightly constrained, and any inflationary scenario that generates enhanced power at small scales must remain consistent with these bounds. In particular, the formation of PBHs is extremely sensitive to the amplitude of primordial curvature perturbations, such that even a moderate enhancement of the power spectrum can lead to their overproduction \cite{Sasaki:2018dmp,Germani:2017bcs,Ballesteros:2017fsr}. This makes PBHs a powerful probe of inflationary dynamics on scales far smaller than those accessible through cosmic microwave background observations.

In this work, we investigate the inflationary dynamics and the evolution of primordial perturbations within two representative models of modified teleparallel gravity, namely the power-law and exponential scenarios. In both cases, we consider a string-inspired fiber inflation potential capable of generating a transient USR phase, leading to a significant enhancement of scalar perturbations at small scales. By numerically solving the full system of background and perturbation equations, we quantify how the modified torsional sector affects key inflationary observables, including the scalar power spectrum and the tensor-to-scalar ratio. Furthermore, we analyze the resulting amplification of fluctuations and its implications for PBH formation, assessing the viability of these scenarios in light of current observational constraints.

\section{A Very Short Introduction To Teleparallel Gravity}\label{Intro_TG}

In TG, which can be formulated as a gauge theory of the translation group \cite{Aldrovandi-Pereira-book, JGPereira2, Arcos:2005ec}, gravitation is encoded in torsion rather than curvature \cite{Aldrovandi:2013wha, Pereira:2019woq}. The fundamental dynamical variable is the tetrad field $e^A_{\,\,\mu}$, which relates the spacetime metric to the Minkowski metric in the tangent space through
\be
g_{\mu \nu}=\eta_{A B} e^{A}_{~\mu} e^{B}_{~\nu} \,,
\ee where $\eta _{AB}^{}=\text{diag}\,(-1,1,1,1)$. 
The geometric structure of TG is defined by a curvature-free, metric-compatible connection, known as the Weitzenböck connection. In the covariant formulation, this connection is constructed from the tetrad and the spin connection \cite{Krssak:2018ywd}, which can be written as
\be
\omega^{A}_{~B \mu}=\Lambda^{A}_{~D}(x) \partial_{\mu}{\Lambda_{B}^{~D}(x)} \,,
\label{spin_TG}
\ee where $\Lambda^{A}_{~D}(x)$ denotes a local Lorentz transformation. By construction, this connection has vanishing curvature,
\be
R^{A}_{~B \mu\nu}=\partial_{\mu} \omega^{A}_{~B\nu}-\partial_{\nu}{\omega^{A}_{~B \mu}}+\omega^{A}_{~C \mu} \omega^{C}_{~B \nu}-\omega^{A}_{~C \nu} \omega^{C}_{~B \mu}=0  \,,
\ee 
and is therefore referred to as a flat or purely inertial connection. Nevertheless, it possesses a non-vanishing torsion tensor, which encodes the gravitational degrees of freedom, defined as
\be
T^{A}_{~~\mu \nu}=\partial_{\mu}e^{A}_{~\nu} -\partial_{\nu}e^{A}_{~\mu}+\omega^{A}_{~B\mu}\,e^{B}_{~\nu}
 -\omega^{A}_{~B\nu}\,e^{B}_{~\mu} \,.
\ee
A spacetime connection can be defined from the tetrad and spin connection as
\be
\Gamma^{\rho}_{~~\nu \mu}=e_{A}^{~\rho}\partial_{\mu}e^{A}_{~\nu}+e_{A}^{~\rho}\omega^{A}_{~B \mu} e^{B}_{~\nu} \,,
\ee
which corresponds to the Weitzenb\"ock connection. This connection is related to the Levi-Civita connection $\bar{\Gamma}^{\rho}_{~~\nu \mu}$ 
of GR through the contortion tensor,
\be
\Gamma^{\rho}_{~~\nu \mu}=\bar{\Gamma}^{\rho}_{~~\nu \mu}+K^{\rho}_{~~\nu \mu} \,,
\label{RelGamma}
\ee
where
\begin{equation}  \label{Contortion}
 K^{\rho}_{~~\nu\mu}= \frac{1}{2}\left(T^{~\rho}_{\nu~\mu}
 +T^{~\rho}_{\mu~\nu}-T^{\rho}_{~~\nu\mu}\right) \,.
\end{equation}
The dynamics of TG is not uniquely fixed by the geometric construction alone. In general, one can construct different gravitational actions from scalar quantities built out of the torsion tensor. A particularly important choice is obtained by considering the torsion scalar $T$, defined as
\be
T= S_{\rho}^{~~\mu\nu}\,T^{\rho}_{~~\mu\nu},
\label{ScalarT}
 \ee
where
\begin{equation} \label{Superpotential}
 S_{\rho}^{~~\mu\nu}=\frac{1}{2}\left(K^{\mu\nu}_{~~~\rho}+\delta^{\mu}_{~\rho} \,T^{\theta\nu}_{~~~\theta}-\delta^{\nu}_{~\rho}\,T^{\theta\mu}_{~~~\theta}\right)\,,
\end{equation}
is the superpotential tensor. When the gravitational action is taken to be linear in 
$T$
\be
S=-\frac{1}{2 \kappa^2} \int{d^{4}x e ~T},
\ee
with $\kappa^2 = 8 \pi G$ and $e=\det{(e^{A}_{~\mu})}=\sqrt{-g}$, one recovers the Teleparallel Equivalent of General Relativity \cite{Aldrovandi:2013wha}. This specific choice is distinguished by the fact that the torsion scalar and the Ricci scalar of GR differ only by a total derivative
\be
T=-\bar{R}+2 e^{-1} \partial_{\mu}(e T^{\nu \mu}_{~~~\nu})  \,,
\label{Equiv} 
\ee
which ensures that TG and GR are dynamically equivalent at the level of the field equations. It is important to emphasize that, despite this equivalence at the level of the field equations, the underlying geometric interpretation of gravity in TG differs fundamentally from that of GR.

This equivalence provides a natural starting point for the construction of modified theories of gravity. In particular, extensions based on nonlinear functions of the torsion scalar, such as $f(T)$ gravity \cite{Bengochea:2008gz, Linder:2010py}, give rise to genuinely new theories that are not equivalent to their curvature-based counterparts. In addition, models involving scalar fields nonminimally coupled to torsion have been extensively studied in the context of dark energy and inflation \cite{Geng:2011aj, Otalora:2013tba, Otalora:2013dsa, Otalora:2014aoa, Skugoreva:2014ena}. These frameworks exhibit a rich phenomenology, motivating their application to both early- and late-time cosmology \cite{Cai:2015emx}.

\section{Scalar -- Torsion Gravity }

In analogy to curvature-based modified gravity models \cite{Fujii:2003pa,faraoni2004cosmology,Tsujikawa:2008uc,Alimohammadi:2009yt}, one can generalize TG by promoting the gravitational action as a function of both the torsion scalar and a scalar field. The resulting scalar–torsion theory is described by an action of the form \cite{Hohmann:2018rwf,Gonzalez-Espinoza:2020azh}
\begin{equation}
 S=\int d^{4}x\,e\,\left[ f(T,\phi)+ P(\phi)X \right] \,,
\label{action1}
\end{equation}
where $f(T, \phi)$ is an arbitrary function of the torsion scalar $T$ and the scalar field $\phi$, while $X=-\partial_ {\mu}{\phi}\partial^{\mu}{\phi}/2$ denotes the kinetic term. The function $P(\phi)$ allows for a nontrivial kinetic coupling.

This action encompasses a broad class of torsion-based theories, including $f(T)$ gravity, minimally coupled scalar field models, and nonminimally coupled scalar–torsion theories. In particular, for the choice $f(T,\phi)=-M_{pl}^2 T/2-V(\phi)$, one recovers TG (equivalently GR) coupled to a canonical scalar field with potential
$V(\phi)$ \cite{MukhanovBook}. Here, $M_{pl}=1/(8\pi G)^{1/2}$ denotes the reduced Planck mass.
Although the TG is dynamically equivalent to GR, the generalized action above defines a genuinely distinct class of modified gravity theories, with no direct curvature-based analogue and a rich phenomenology \cite{Cai:2015emx}.

Varying the action with respect to the tetrad field leads to modified gravitational field equations of the form
\bea
 && f_{,T} G_{\mu \nu}+S_{\mu \nu}{}^{\rho} \partial_{\rho} f_{,T}+\frac{1}{4}g_{\mu \nu}\left(f-T f_{,T}\right)+\frac{P}{4}\left(g_{\mu \nu} X+\partial_{\mu}\phi \partial_{\nu}\phi\right)=0 \,,
\label{FieldEquations}
\eea
where $f_{,T} \equiv \partial f / \partial T$ and $S_{\mu\nu}{}^{\rho}$ is the superpotential tensor, which is linear in the torsion tensor. The tensor $G^{\mu}_{~\nu}$ denotes the Einstein tensor constructed from the tetrad variables.

It is important to note that, in general, this class of theories is not invariant under local Lorentz transformations. As a consequence, the field equations are not symmetric, reflecting the presence of additional degrees of freedom beyond those of GR. In particular, the superpotential tensor is not symmetric in its lower indices. These extra degrees of freedom can play a significant role in the evolution of cosmological perturbations and must be properly taken into account in inflationary scenarios \cite{Gonzalez-Espinoza:2020azh}.

\subsection{Inflation in $f(T,\phi)$ gravity }

We consider a homogeneous and isotropic cosmological background described by a spatially flat Friedmann–Lemaître–Robertson–Walker (FLRW) spacetime, with line element
\begin{equation}
    ds^2= -dt^2 + a(t)^2\delta_{ij}dx^idx^j \,,
\end{equation}
where $a(t)$ is the scale factor. A suitable choice of tetrad compatible with this geometry is given by the diagonal configuration
\begin{equation}
    e^{A}{}_{\mu}=\text{diag}(1,a(t),a(t),a(t)) \,.
\end{equation}
The Hubble parameter is defined as $H=\dot{a}/a$. For this background, the torsion scalar reduces to $T=-6H^2$, which allows us to express the field equations in terms of the Hubble rate.

Substituting this ansatz into the field equations, we obtain the modified background equations governing the cosmological dynamics,
\begin{align}
    f(T,\phi)-P(\phi)X-2Tf_{,T} &=0  \,,   \\
    f(T,\phi)+P(\phi)X-2Tf_{,T}-4\dot H f_{,T}-4H \dot f_{,T}&=0 \,,    \\
    -P_{,\phi X}-3P(\phi)H\dot \phi - P(\phi)\ddot \phi +f_{,\phi}=0  \,,
\end{align}
where $X=- \dot{\phi}^2/2$. A comma denotes partial derivatives, for instance $f_{,T}=\partial f / \partial T$ and $f_{, \phi} = \partial f / \partial \phi$, while a dot represents derivatives with respect to cosmic time.

To characterize the inflationary dynamics, we introduce the slow-roll parameters
\begin{equation}
    \epsilon\equiv-\frac{\dot H}{H^2} \,, \quad \quad  \delta P_X \equiv -\frac{P(\phi)X}{2H^2 f_{,T}} \,, \quad \quad  \delta f_{,T}\equiv \frac{\dot f_{,T}}{f_{,T}H} \,.
\end{equation}
The first slow-roll parameter can then be written as 
\begin{equation}
    \epsilon = \delta_{PX} + \delta_{f,T} \,.
\end{equation}
The quantity $\delta_{f_{,T}}$ encodes the deviations from TG and can be further decomposed as 
\begin{equation}
    \delta_{f_{,T}} = \delta_{f\dot H }+\delta_{fX} \,,
\end{equation}
where
\begin{equation}
    \delta_{f\dot H } \equiv \frac{f_{,TT}\dot T }{H f_{,T}} \,,  \quad \quad  \delta_{fX} \equiv \frac{f_{,T\phi}\dot \phi }{H f_{,T}} \,.
\end{equation}
By combining the background equations, these quantities can be expressed in terms of a dimensionless parameter that characterizes deviations from GR. In particular, one finds 
\begin{align}
    \delta_{f\dot H } &= -\frac{2\mu}{1+2\mu}(\delta_{PX} + \delta_{fX}) \,, \\
    \delta_{f_{,T}} &= \frac{1}{1+2\mu}(\delta_{fX}-2\mu \delta_{PX}) \,,  \\
    \epsilon &= \frac{1}{1+2\mu}(\delta_{PX}+\delta_{fX}) \,,
\end{align}
where the parameter
\begin{equation}
\mu = f_{,TT}/f_{,T} \,,
\end{equation}
quantifies the deviation from TG.
These expressions explicitly show how the torsional sector modifies the inflationary dynamics. In particular, the parameter $\mu$ controls the departure from GR, leading to a rescaling of the slow-roll parameter $\epsilon$. In the limit $\mu \rightarrow 0$, the standard results of canonical single-field inflation are recovered.

\subsection{Scalar Perturbations}
To study primordial fluctuations, we begin with the Arnowitt–Deser–Misner (ADM) decomposition of the tetrad field. In this formalism, the tetrad components are written as
\begin{align}
e^{0}{}_\mu &=(\mathcal{N},0)\,, \quad e^{a}{}_\mu=(\mathcal{N}^a,h^{a}{}_i)\,,
\end{align}
and their inverses
\begin{align}
e_{0}{}^{\mu} &=(1/\mathcal{N},-\mathcal{N}^i/\mathcal{N}) \,, \quad e_{a}{}^\mu=(0,h_{a}{}^i)\,.
\end{align}
Here, $\mathcal{N}$ is the lapse function and $\mathcal{N}^i=h_a{}^i \mathcal{N}^a$ is the shift vector, while $h^a{}_i$ denotes the induced spatial tetrad satisfying the orthonormality condition $h^a{}_jh_a{}^i=\delta^i_j$.

To analyze scalar perturbations, we work in the comoving gauge defined by $\delta \phi = 0$. In this gauge, the ADM variables are parametrized as
\begin{equation}
    N = 1+\alpha \,, \quad N^a=a^{-1}e^{-\mathcal{R}}\delta^a{}_i \partial^i\psi \,, \quad h^a{}_i = a e^{\mathcal{R}}\delta ^a{}_j\delta ^j{}_i \,,
\end{equation}
where $\mathcal{R}$ is the comoving curvature perturbation. With this choice, the perturbed metric takes the form
\begin{equation}
    ds^2 = - \left[(1+ \alpha)^2-a^{-2}e^{-2\mathcal{R}} (\partial \psi)^2\right]dt^2+2\partial_i\psi dtdx^i+a^2 e^{2\mathcal{R}}\delta_{ij}dx^idx^j \,.
\end{equation}
Expanding the action up to second order in scalar perturbations, one obtains the quadratic action \cite{Gonzalez-Espinoza:2020azh}
\begin{equation}
    S^{(2)}=\frac{1}{2}\int d\tau dx^3[(v')^2-(\partial v)^2-M^2v^2] \,,
\end{equation}
where primes denote derivatives with respect to conformal time $\tau$.

The comoving curvature perturbation $\mathcal{R}$ is related to the Mukhanov–Sasaki variable through 
\begin{equation}
    v \equiv z \mathcal{R} \,, \quad z^2 = 2a^2Q_s \,,
\end{equation}
where $Q_s = PX/H^2$.

In modified teleparallel gravity, the dynamics of perturbations is affected by an effective mass term given by
\begin{equation}
    M^2 \equiv a^2m^2-\frac{z''}{z} \,,
\end{equation}
where $m^2 = 3H^2 \eta_\mathcal{R}$.
The parameter $\eta_\mathcal{R}$ encodes the effects of torsional modifications and can be written as
\begin{equation}
    \eta_\mathcal{R} = \frac{m^2}{3H^2} =\delta_{f_{,T}}\left [1 + \left(1+\frac{\delta_{fX}}{\delta_{PX}}\right)\frac{\delta_f{_{,T}}}{\delta_{f\dot H}} \right] \,.
\end{equation}
Finally, by varying the quadratic action, one obtains the Mukhanov–Sasaki equation in Fourier space
\begin{equation}
    v''_k+(k^2+M^2)v_k=0 \,. \label{mukhanov}
\end{equation}

\subsection{Slow-Roll Dynamics in Modified Teleparallel Gravity}

Assuming that the standard slow-roll regime takes place prior to the onset of the USR phase, the background dynamics can be treated within the slow-roll approximation. In this regime, the kinetic energy of the scalar field is subdominant and the time variation of the Hubble parameter is small, allowing a significant simplification of the background equations.

Under these assumptions, the modified Friedmann and Klein–Gordon equations reduce to 
\begin{align}
    f(T,\phi) = 2T f_{,T}  \,, \\
    3P(\phi)H \dot \phi=f_{,\phi}\,.
\end{align}
These relations generalize the standard slow-roll conditions in GR, with the important difference that the function $f(T, \phi)$ and its derivatives encode the effects of the torsional sector. In particular, the evolution of the scalar field is directly influenced by the coupling to torsion through the term $f_{,\phi}$.

Within this approximation, the scalar power spectrum evaluated at horizon crossing can be written as
\begin{equation}
\mathcal{P}_\mathcal{R} \equiv \frac{k^3}{2 \pi^2}|\mathcal{R}_k|^2 \simeq\frac{H_k^2}{8 \pi Q_{s}}\left[ 1+ 2 \eta_R  \ln \left(\frac{k}{aH}\right)\right ] \,,
\end{equation}
and the tensor power spectrum is given by \cite{Gonzalez-Espinoza:2020azh}
\begin{equation}
    \mathcal{P}_T =\frac{H^2}{2\pi^2Q_{Tk}} \,.
\end{equation}
Using the slow-roll approximation, we can numerically obtain the initial conditions for the parameters $H(N)$, $\phi(N)$ and $\phi'(N)$ in terms of the number of e-folds by considering the end of this regime at the start of the USR phase. This allow us to obtain the values of the parameters at the horizon crossing.

\section{Inflation in Modified Teleparallel Gravity Models}

In this work, we focus on scalar–torsion gravity models of the form $f(T,\phi) = f(T) -V(\phi)$, neglecting nonminimal couplings between the scalar field and torsion, i.e., $f_{,T\phi}=0$. We consider a canonical kinetic term by setting $P(\phi)=1$. Within this framework, we study two representative models for $f(T)$: the power-law and exponential cases, treating them as effective modifications of teleparallel gravity \cite{Tzerefos:2023mpe}. The scalar potential $V(\phi)$ is chosen to be of the fibre inflation type, allowing for the realization of a transient USR phase.

\subsection{Fibre Inflationary Potential}

Fibre inflation models arise in the context of type IIB string theory compactifications with fluxes and moduli stabilization. In these constructions, inflation is driven by a Kähler modulus associated with a fibred Calabi–Yau manifold, typically involving a K3 fibration over a $\mathbb{P}^1$ base \cite{Cicoli:2008gp, Cicoli:2018asa}. The overall volume of the compact space can be written as 
\begin{equation}
    \mathcal{V}= t_P \tau_{K3}-\tau_{dP}^{3/2}\,,
\end{equation}
where $\tau_{K3}$ corresponds to the fibre modulus and plays the role of the inflaton, while $\tau_{dP}$ denotes a blow-up mode responsible for moduli stabilization.

The inflaton field is related to the K\"ahler modulus $\tau_{K3}$ through the field redefinition
\begin{equation}
    \tau_{\text{K}3}= e^{\frac{2}{\sqrt{3}}\phi} = \langle \tau_{\text{K}3}\rangle e^{\frac{2}{\sqrt{3}}\hat\phi}\,,
\end{equation}
such that the canonically normalized scalar field can be expressed as
\begin{equation}
\phi = \frac{\sqrt{3}}{2} \ln \langle\tau_{\text{K}3}\rangle + \hat\phi  \,.
\end{equation}
This parameterization allows one to describe the inflationary dynamics in terms of the displacement $\hat{\phi}$ around the minimum of the potential.

Fibre inflation models are characterized by scalar potentials generated by a combination of string loop corrections and higher-derivative effects, leading to a plateau-like structure suitable for slow-roll inflation \cite{Cicoli:2008gp}. Importantly, these potentials can naturally exhibit near-inflection points, which can trigger a transient USR phase and lead to a significant enhancement of scalar perturbations at small scales.

Since our aim is to study the production of PBHs through such an enhancement, we consider a generalized form of the fibre inflation potential given by \cite{Cicoli:2018asa}
\begin{equation}
    V_{inf} = \frac{W_0^2}{\mathcal{V}^3}\left[  - \frac{C_W}{\sqrt{\tau_{K3}}} +\frac{A_W}{\sqrt{\tau_{K3}} - B_W}-\frac{\tau_{K3}}{\mathcal{V}} \left( \frac{G_W}{1+R_W \frac{\tau_{K3}^{3/2}}{\mathcal{V}}}\right)\right] \,.
\end{equation}
This effective potential captures the leading contributions relevant for inflation, including corrections that can generate a near-inflection point. Such features are crucial for inducing a temporary violation of the slow-roll conditions, which in turn leads to a rapid growth of curvature perturbations and enhances the primordial power spectrum.

The parameters of the model are chosen such that they correspond to realistic compactifications and yield viable inflationary dynamics. In particular, we consider several benchmark configurations, denoted by $V_1$-$V_4$, whose parameter values are summarized in Table~\ref{tab:Potential}. These setups are selected to produce a significant enhancement of the scalar power spectrum while remaining consistent with large-scale cosmological constraints.

\begin{figure}[H]
    \centering
    \includegraphics[width=0.7\linewidth]{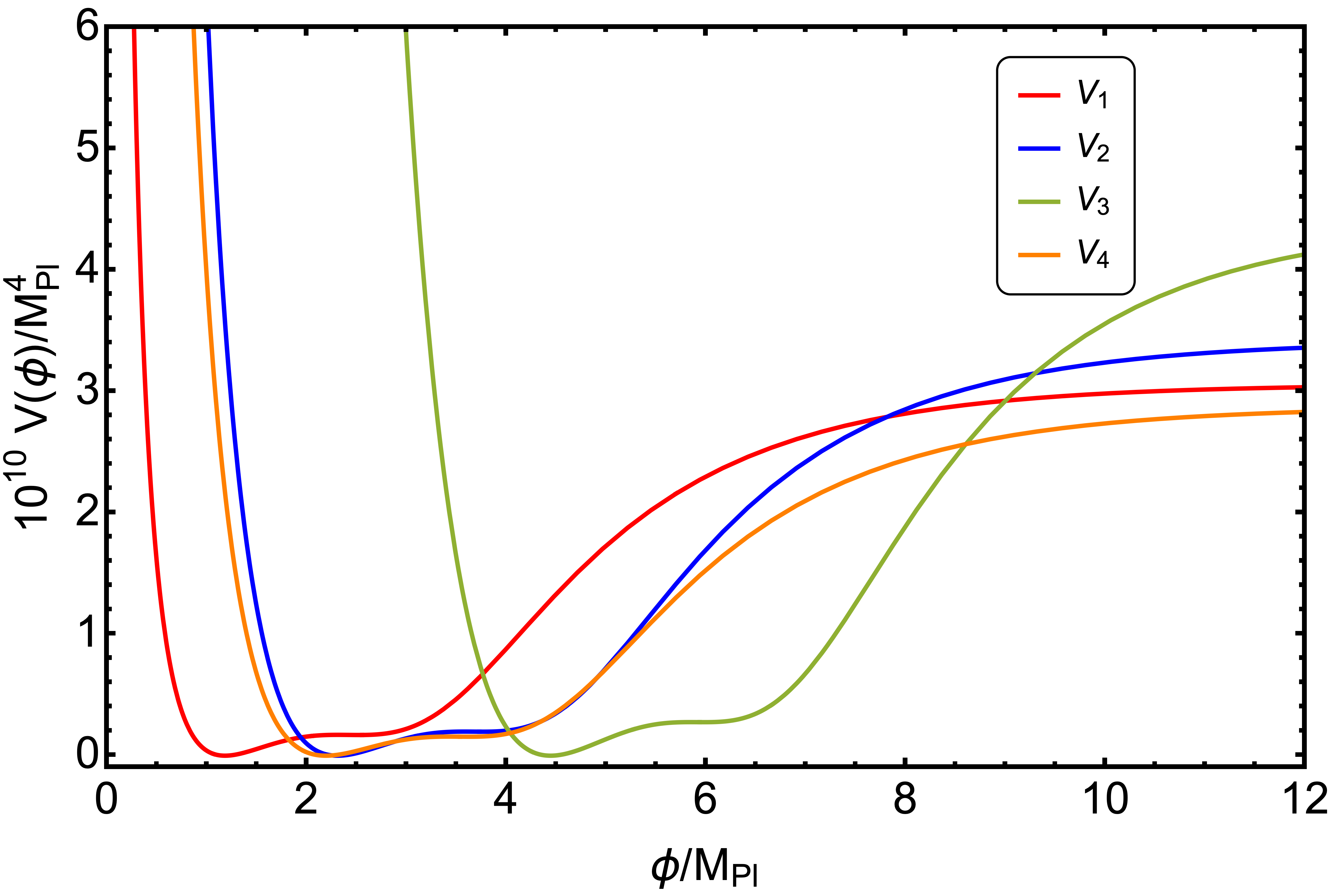}
    \caption{Fibre inflation potentials corresponding to the configurations listed in Table \ref{tab:Potential}. The potentials exhibit a plateau region followed by a near-inflection point, which can trigger a transient ultra slow-roll phase.}
    \label{fig:PotentialPlot}
\end{figure}

In Fig.~\ref{fig:PotentialPlot}, we display the corresponding scalar potentials for these configurations. As can be seen, the potentials exhibit a characteristic plateau at large field values, ensuring a standard slow-roll phase, followed by a near-inflection region. This feature is responsible for triggering the USR phase, during which the slow-roll parameter $\epsilon$ becomes strongly suppressed. As a consequence, the curvature perturbations grow on superhorizon scales, providing the necessary conditions for efficient primordial black hole production.
\begin{table}[H]
\centering
\resizebox{1.0\textwidth}{!}{%
\begin{tabular}{|c|c|c|c|c|c|c|c|c|}
\hline
\textbf{Model} & $\mathbf{W_0}$ & $\mathbf{C_W}$ & $\mathbf{A_W}$ & $\mathbf{B_W}$ & $\mathbf{G_W/\mathcal{V}}$ & $\mathbf{R_W/\mathcal{V}}$ & $\mathbf{\mathcal{V}}$ & $\mathbf{\langle \tau_{K3}\rangle}$ \\
\hline
\textbf{$V_1$} & $1.04 \times 10^{-1}$ & $1/10$ & $2/100$ & $1$ & $1.303386 \times 10^{-5}$ & $6.58724\times 10^{-3}$ & $107.3$ & $3.88738$\\
\hline
\textbf{$V_2$} & $9.4690334348$ & $4/100$ & $2/100$ & $1$ & $3.080547\times 10^{-5}$ & $7.071067\times 10^{-4}$ & $1000$ & $14.2957$ 
\\
\hline
\textbf{$V_3$} & $1.85 \times 10^{4}$ & $1.978 \times 10^{-2}$ & $1.65 \times 10^{-2}$ & $1.01$ & $9.257715\times 10^{-8}$ & $1.414\times 10^{-5}$ & $5 \times 10^{4}$ & $168.033$ 
\\
\hline
\textbf{$V_4$} & $1.60$ & $17/100$ & $8/100$ & $1$ & $1.8001844 \times 10^{-4}$ & $9.1079272 \times 10^{-4}$ & $550$ & $12.4865$ \\
\hline
\end{tabular}
}
\caption{Fibre inflation potential parameter sets leading to an enhancement of the primordial power spectrum \cite{Cicoli:2018asa}.}
\label{tab:Potential}
\end{table}

\subsection{Power-Law Model}

We begin with the power-law (PL) form of $f(T)$ \cite{Bengochea:2008gz}, given by
\begin{equation}
    f(T) = -\frac{M_{Pl}^2}{2}\left(T+ \alpha T^{\beta} \right)\,,
\end{equation}
where $\alpha$ and $\beta$ are free parameters. For the purpose of studying small deviations from TG we fix $\beta = 2$ to study a specific model, and the parameter $\alpha$ will act as a correction term. \\
Substituting this form of $f(T)$ into the gravitational and Klein-Gordon equations, and expressing derivatives with respect to the number of e-folds $N$, we obtain the following background equations
\begin{equation}
\frac{ M_{pl}^2T}{2} + \frac{T^\beta M_{pl}^2 \, \alpha\,  (2\beta - 1)}{2} - \frac{H^2 \phi'^2}{2} - V = 0 \,, 
\label{background1}
\end{equation}
\begin{equation}
\frac{ M_{pl}^2 T}{2}+ \frac{M_{pl}^2 T^{\beta} \alpha(2\beta -1)}{2} - 2M_{pl}^2\, \epsilon\,  H^2- 2M_{pl}^2 \, \epsilon\,  H^2 \alpha\,  \beta\,  (2\beta-1)T^{\beta-1} +  \frac{H^2 \phi'^2}{2}-V =0\,,
\label{background2}
\end{equation}
\begin{equation}
\phi'' + (3-\epsilon)\phi' + V_{,\phi}/H^2 = 0    \,,
\end{equation}
where primes denote derivatives with respect to $N$.
Since our primary goal is to study the production of primordial black holes, it is necessary to solve the full system of background equations numerically, without relying on the slow-roll approximation. This is particularly important in the presence of a USR phase, where the slow-roll conditions are temporarily violated.

From Eqs. \eqref{background1} and \eqref{background2} we obtain the following expression for the first slow-roll parameter in terms of the number of e-folds
\begin{equation}
    \epsilon  = \frac{\phi'^2 }{2M_{Pl}^2\left(1+\alpha \beta (2\beta-1) T^{2\beta-1} \right)} \,.
\end{equation}
The denominator introduces a correction to the standard GR result, which can either enhance or suppress the value of $\epsilon$ depending on the sign and magnitude of $\alpha$. By taking the derivative of the first slow-roll parameter with respect to the e-fold number, we obtain the second slow-roll parameter
\begin{align}
\notag \eta=&- \frac{
6^{-1+\beta}\alpha\beta(\beta-1)(2\beta-1)
H \left(H^2\right)^{\beta-2} H' \left(\phi'\right)^2
}{M_{Pl}^2\left[ 1 + 6^{-1+\beta}\alpha\beta(2\beta-1)
\left(H^2\right)^{\beta-1}
\right]^2} \\
&+ \frac{\phi'\phi''}{M_{Pl}^2
\left[1 + 6^{\beta-1}\alpha\beta(2\beta-1)
\left(H^2\right)^{\beta-1}\right]} \,.
\end{align}
In Fig \ref{fig:USR-PL-EXP-Model} we present the inflationary dynamics of the PL model for different values of the correction parameter $\alpha$, along with different configurations of the fibre inflation potential. The values of $\alpha$ considered  has an impact on the inflationary evolution, in particular, as it can be seen in the figure, increasing $\alpha$ leads to a reduction in the total number of e-folds and also the maximum decrease of the first slow-roll parameter $\epsilon$ during the USR phase, which directly affects the enhancement in the primordial power spectrum.

\subsection{Exponential Model}

In this subsection, we consider an exponential form of $f(T)$, motivated by analogous constructions in exponential $f(R)$ gravity \cite{Linder:2009jz}. The model is defined as
\begin{equation}
    f(T) = -\frac{M_{Pl}^2}{2}\left[T + \alpha T_0\left( 1-e^{-T/(\beta T_0)}\right) \right] \,,
\end{equation}
where $\alpha$ and $\beta$ are dimensionless parameters controlling the strength and scale of the torsional corrections \cite{Nesseris:2013jea}, and sets the characteristic scale of the theory. In what follows, we fix $\beta = 2$, while $\alpha$ parametrizes deviations from TG.

Substituting this form into the field equations and expressing all quantities in terms of the number of e-folds $N$, we obtain the modified Friedmann equations
\begin{equation}
    3H^2M_{pl}^2-3H_0^2M_{pl}^2\alpha  \left( 1-e^{-T/(\beta T_0)}\right)+ \frac{6H^2 M_{pl}^2\alpha}{\beta}e^{\frac{-H^2}{\beta H_0^2}}  - \frac{\phi'^2H^2}{2}-V= 0  \,,
\end{equation}
\begin{equation}
\begin{aligned}
   3H^2M_{pl}^2 -3H_0^2M_{pl}^2\alpha  \left( 1-e^{-T/(\beta T_0)}\right)&+\frac{6H^2 M_{pl}^2\alpha}{\beta}e^{\frac{-H^2}{\beta H_0^2}}-2\epsilon H^2M_{pl}^2\left(1+\frac{1}{\beta}e ^{\frac{-H^2}{\beta H_{0}^2}}\right) \\
   &+\frac{4H^4M_{pl}^2\alpha \epsilon e^{\frac{-H^2}{\beta H_0^2}}}{\beta^2 H_0^2 } - \frac{\phi'^2H^2}{2}-V = 0 \,,
\end{aligned}
\end{equation}
together with the Klein–Gordon equation
\begin{equation}
    \phi'' + (3 - \epsilon)\phi' + \frac{V_{,\phi}}{H^2} = 0  \,.
\end{equation}
As in the PL case, this system must be solved numerically in order to accurately capture the full inflationary dynamics, particularly in the presence of an USR phase.

From the background equations, one obtains the following expression for the first slow-roll parameter: 
\begin{equation}
    \epsilon= -\frac{\phi'^2 \beta^2H_0^2e^{\frac{H^2}{\beta H_0^2}}}{2M_{Pl}^2(2\alpha H^2-H_0^2\alpha\beta-H_0^2 \beta^2 e^{\frac{H^2}{\beta H_0^2}})} \,,
\end{equation}
while the second slow-roll parameter is obtained from its definition from the derivative of the first slow-roll parameter with respect the e-fold number.
\begin{align}
\eta=\frac{e^{\frac{H^2}{H_0^2\beta}} \beta \phi'
\left[HH' \phi'\alpha(-2H^2 + 3H_0^2\beta) + H_0^2\beta
\left(-2H^2\alpha + H_0^2\beta\phi''\left(\alpha + e^{\frac{H^2}{H_0^2\beta}}\beta\right)\right)\right]
}{M_{Pl}^2\left[-2H^2\alpha + H_0^2\beta\left(\alpha + e^{\frac{H^2}{H_0^2\beta}}\beta\right)\right]^2}  \,.
\end{align}

\begin{figure}[H]
    \centering

    \begin{minipage}[t]{0.48\textwidth}
        \centering
        \includegraphics[width=\linewidth]{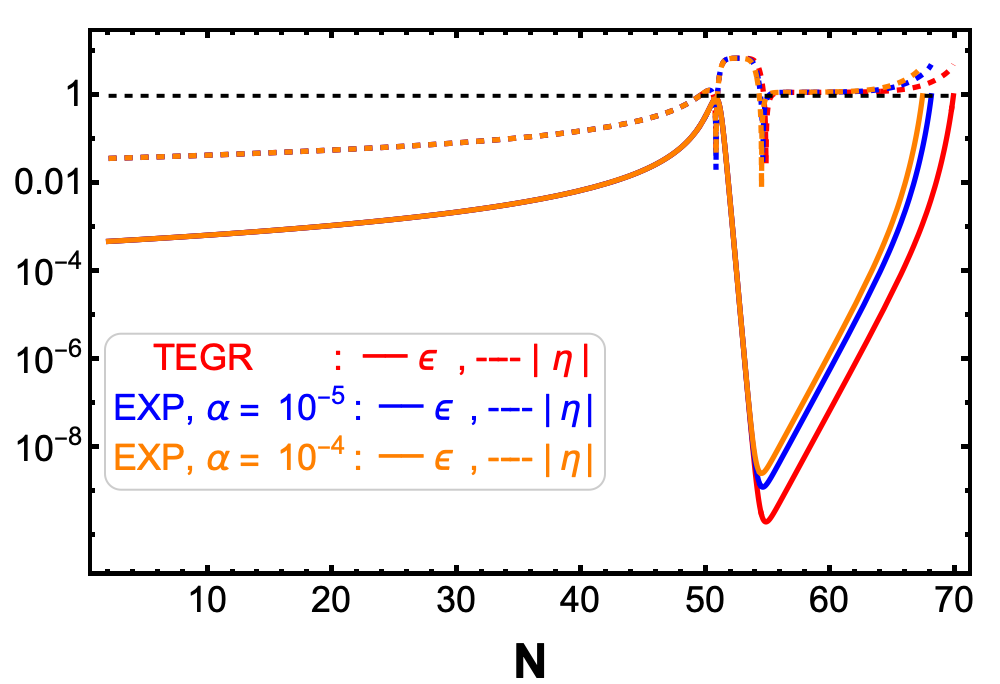}
        \textbf{(a) Configuration $V_2$ Exponential}        
    \end{minipage}
    \hfill
    \begin{minipage}[t]{0.48\textwidth}
        \centering
        \includegraphics[width=\linewidth]{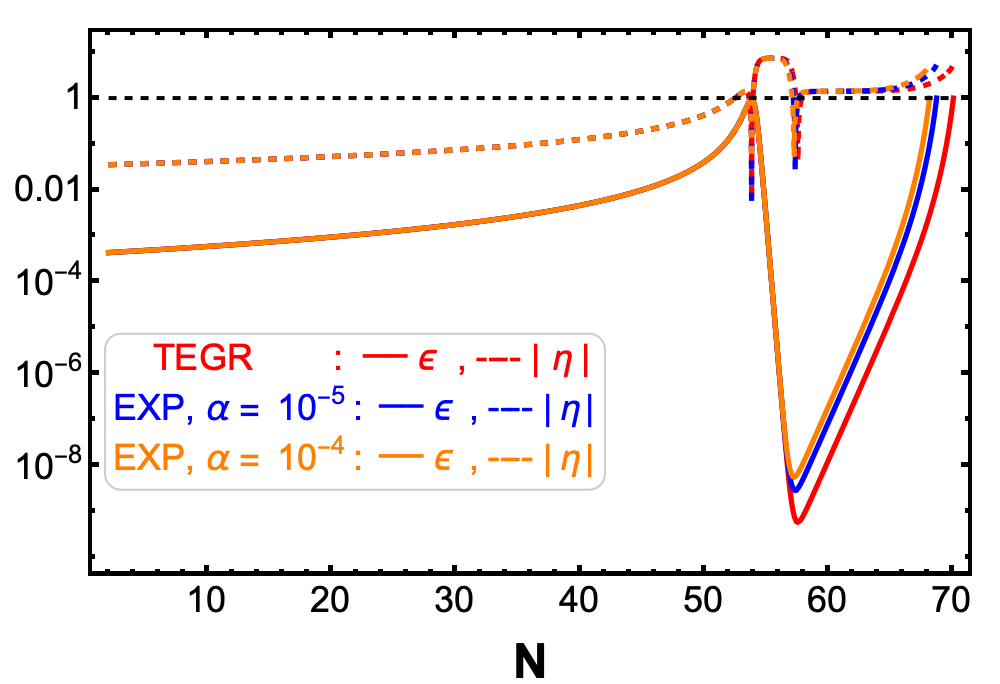}
        \textbf{(b) Configuration $V_4$ Exponential}        

    \end{minipage}
\caption{Inflationary dynamics for the power-law model (top panels) and the exponential model (bottom panels), for different values of the parameter $\alpha$ and for the $V_2$ and $V_4$ configurations of the fibre inflation potential.}
\label{fig:USR-PL-EXP-Model}
\end{figure}

In Fig \ref{fig:USR-PL-EXP-Model} we present the inflationary dynamics for the exponential model considering different values of the parameter $\alpha$ and several configurations of the fibre inflation potential.

\section{Results}

In this section, we present the main results of our analysis of inflationary dynamics and PBH production within modified teleparallel gravity. We begin by examining the background evolution for both the power-law and exponential $f(T)$ models, focusing on the impact of torsional corrections on the slow-roll parameters and the onset of the USR phase. We then compute the evolution of scalar perturbations by numerically solving the Mukhanov–Sasaki equation, allowing us to obtain the primordial power spectrum for the different fibre inflation configurations. Finally, we analyze the implications for PBH formation, including the resulting mass spectrum and abundance, and confront the inflationary predictions with current observational constraints. The results are organized into subsections that progressively connect the background dynamics, perturbation evolution, and PBH phenomenology.

\subsection{Inflationary Dynamics}

By solving numerically the background evolution of the inflationary models by using the initial conditions for $H,\phi$ and $\phi'$ obtained from the breaking of the slow-roll conditions, we can analyse how the deviation term $\alpha$ impacts on this process. We can observe that as this terms grows, the duration of the process and the diminution of the first slow-roll parameter $\epsilon$ during the USR phase is decreased. 
\begin{figure}[H]
    \centering

    \begin{minipage}[t]{0.48\textwidth}
        \centering
        \includegraphics[width=\linewidth]{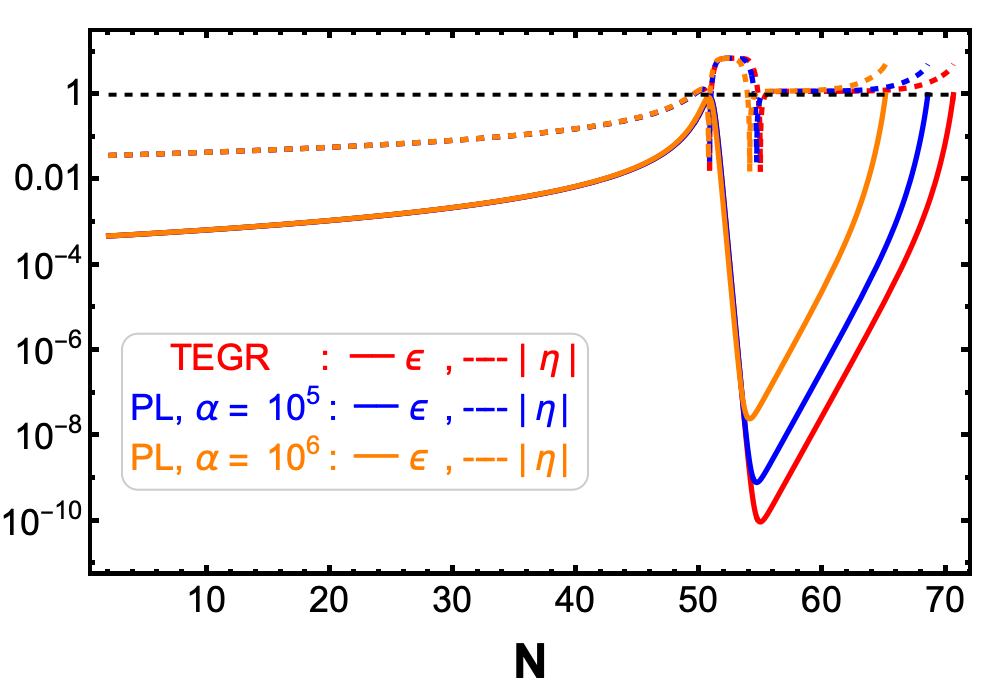}
        \textbf{(a) Configuration $V_2$ Power-Law}        
    \end{minipage}
    \hfill
    \begin{minipage}[t]{0.48\textwidth}
        \centering
        \includegraphics[width=\linewidth]{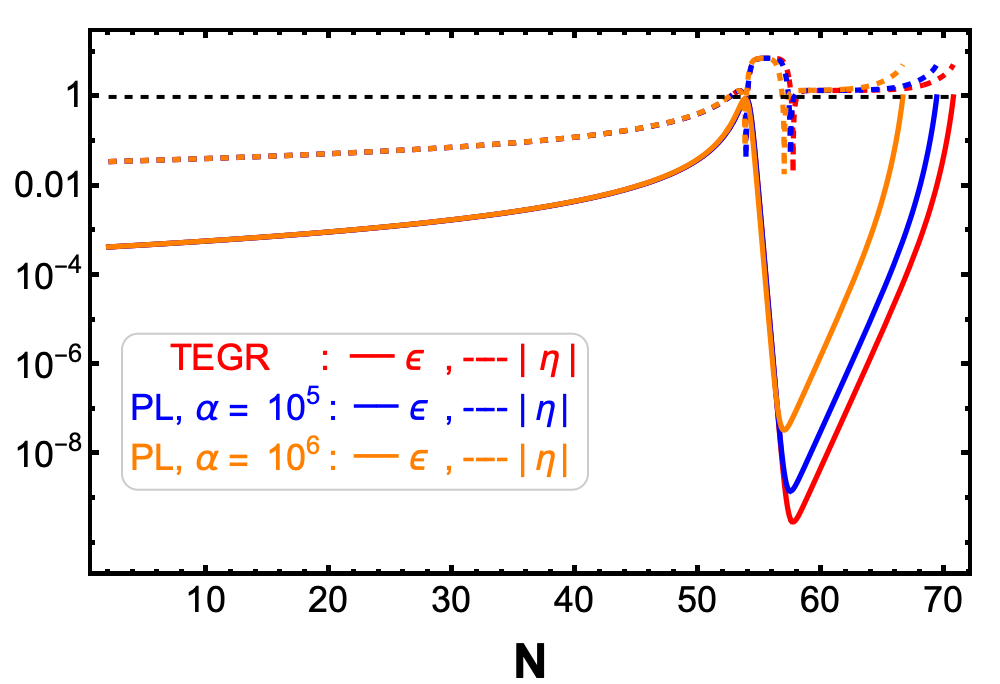}
        \textbf{(b) Configuration $V_4$ Power-Law}        

    \end{minipage}
    
\end{figure}

In the next sections, we consider only the configurations $V_2$ and $V_4$ from the fibre inflationary potential. Configurations $V_1$ and $V_3$ can indeed achieve the peak required in the primordial power spectrum, but the PBHs mass range are severely constrained from Hawking radiation and microlensing constraints respectively.

In the case of the deviation term $\alpha$, we consider values of the order $\mathcal{O}(10^{-5}-10^{-3})$ for the models. This allow us to analyse the effects on the primordial power spectrum enhanced to the required order for the production of PBHs.  

\subsection{Primordial Power Spectrum}
For the production of PBHs it is required the exact evolution of the curvature perturbations beyond the slow-roll approximation. In particular, during the USR phase, the standard slow-roll assumptions break down, and the full Mukhanov–Sasaki equation must be solved numerically.

\begin{figure}[H]
    \centering
    \begin{minipage}[t]{0.48\textwidth}
        \centering
        \hspace{1cm}\textbf{(a) Configuration $V_2$}
        \includegraphics[width=\linewidth]{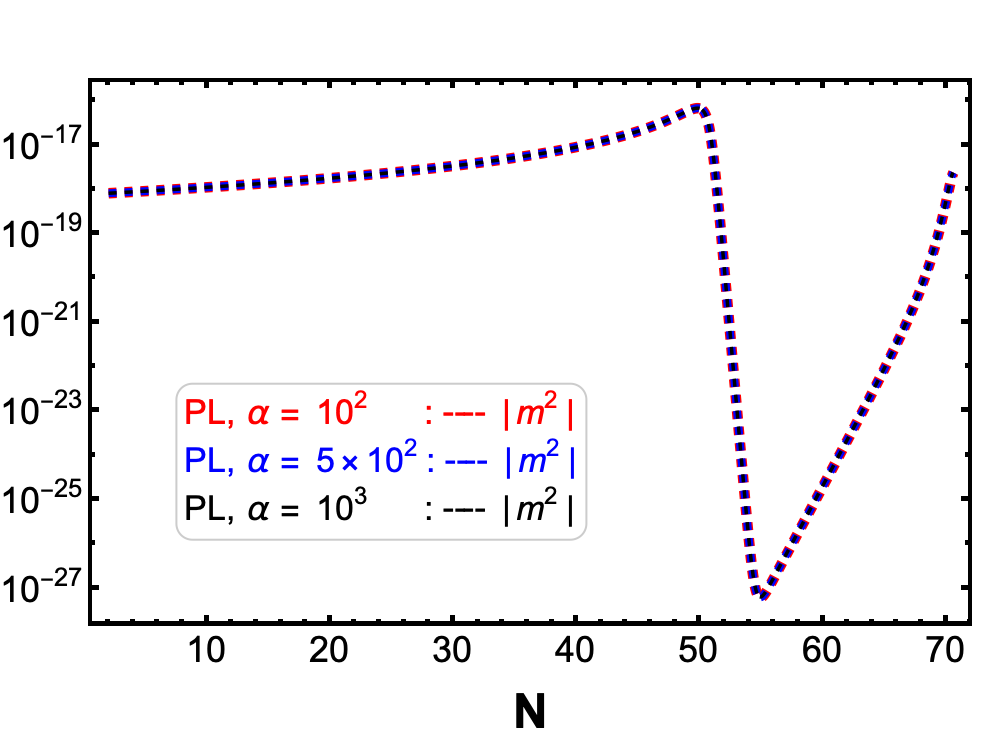}
        
    \end{minipage}
    \hfill
    \begin{minipage}[t]{0.48\textwidth}
        \centering
        \hspace{1cm}\textbf{(b) Configuration $V_4$}
        \includegraphics[width=\linewidth]{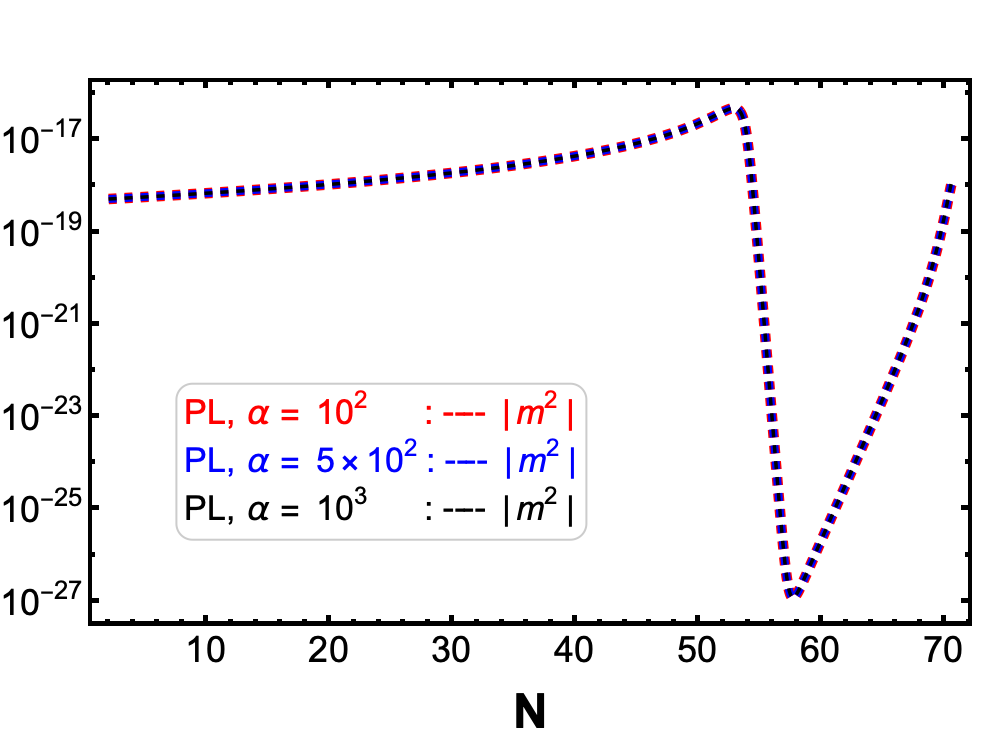}
    \end{minipage}
    \caption{Effective mass term for different fibre inflation configurations in Power-Law model.}
    \label{fig.massPL}

\end{figure}

\begin{figure}[H]
    \centering
    \begin{minipage}[t]{0.48\textwidth}
        \centering
        \includegraphics[width=\linewidth]{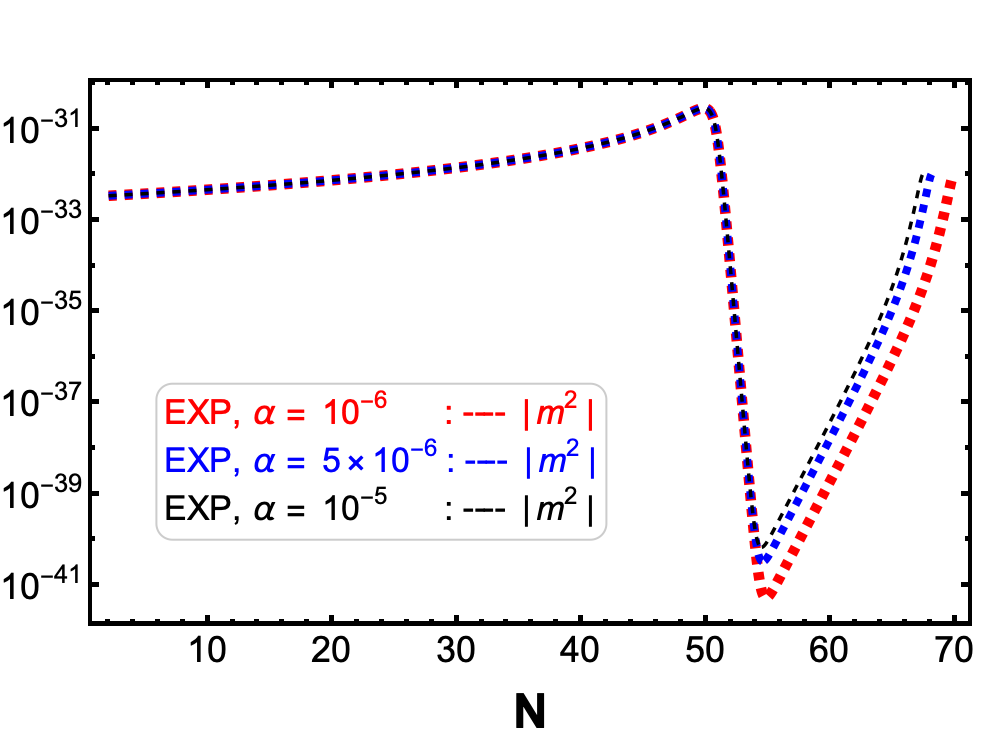}
        \textbf{(a) Configuration $V_2$}        
    \end{minipage}
    \hfill
    \begin{minipage}[t]{0.48\textwidth}
        \centering
        \includegraphics[width=\linewidth]{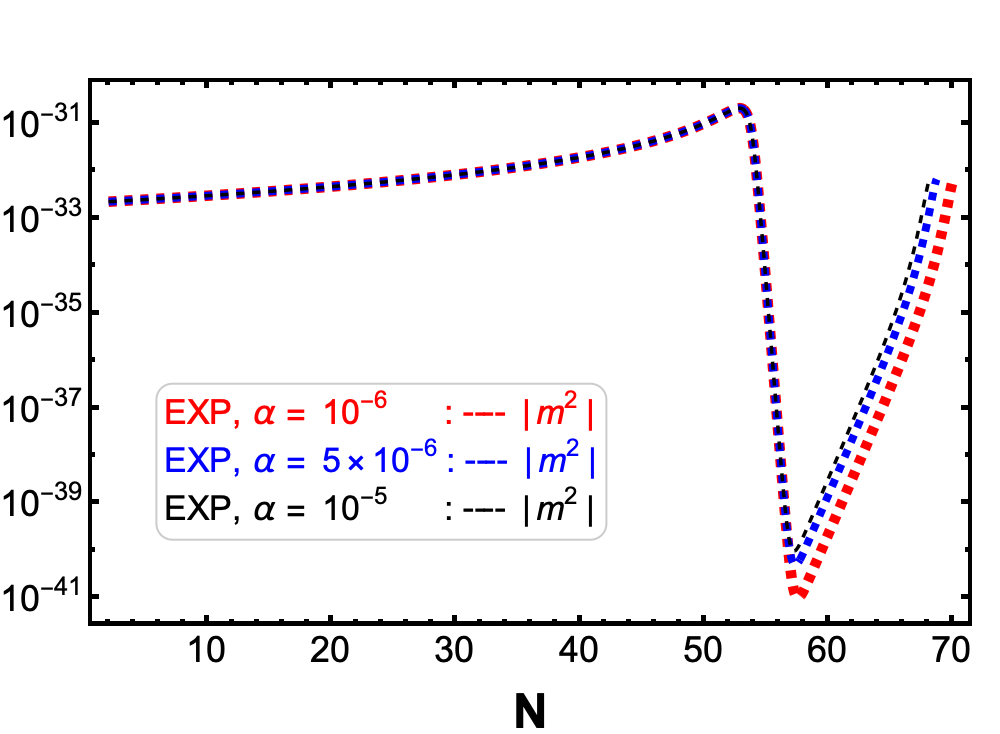}
        \textbf{(b) Configuration $V_4$}             
    \end{minipage}
    \caption{Effective mass term for different fibre inflation configurations in Exponential model.}
        \label{fig.massEXP}

\end{figure}

\begin{figure}[H]
    \centering

    \begin{minipage}[t]{0.48\textwidth}
        \centering
        \hspace{1cm}\textbf{(a) Configuration $V_2$}
        \includegraphics[width=\linewidth]{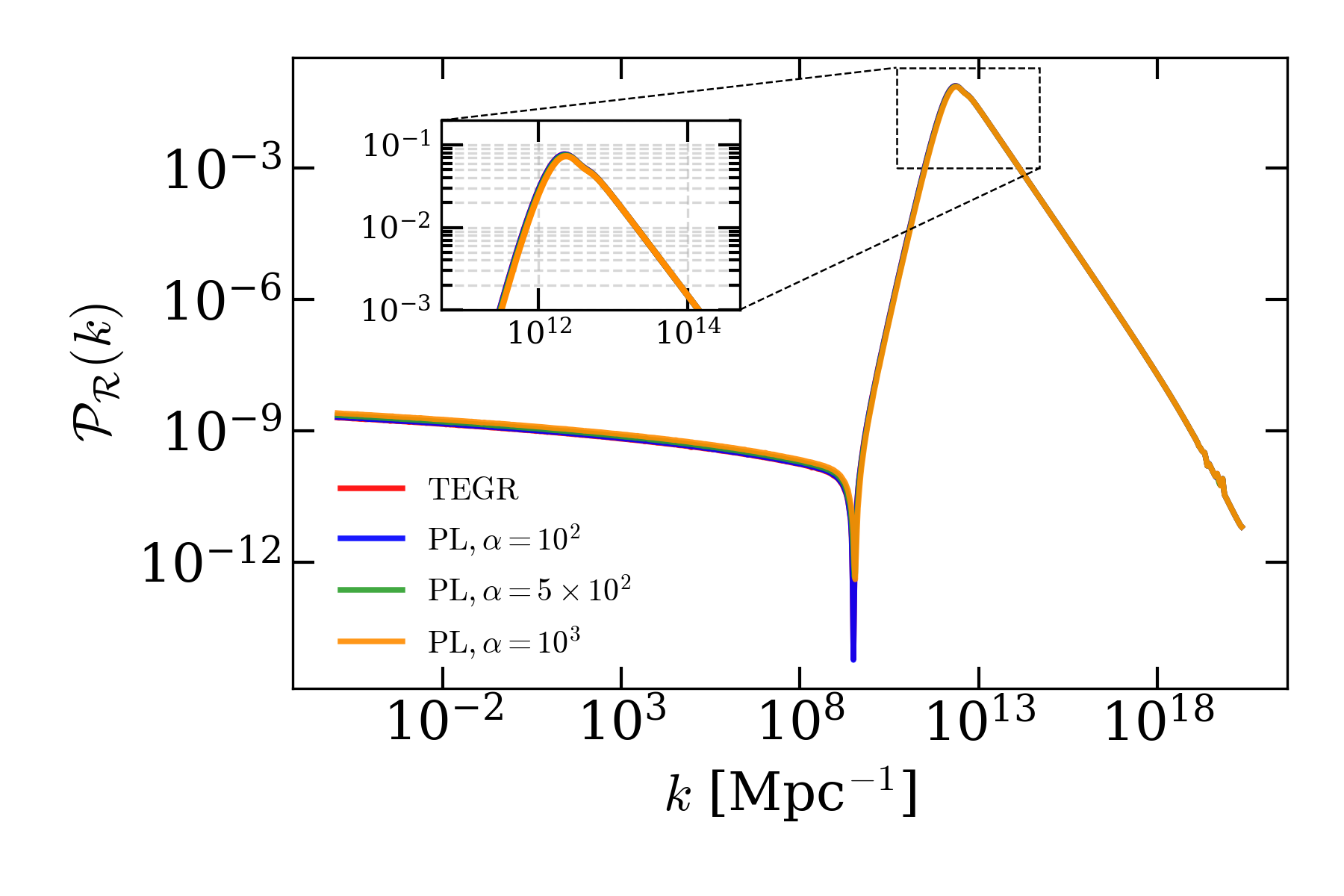}
      
    \end{minipage}
    \hfill
    \begin{minipage}[t]{0.48\textwidth}
        \centering
        \hspace{1cm}\textbf{(b) Configuration $V_4$}
        \includegraphics[width=\linewidth]{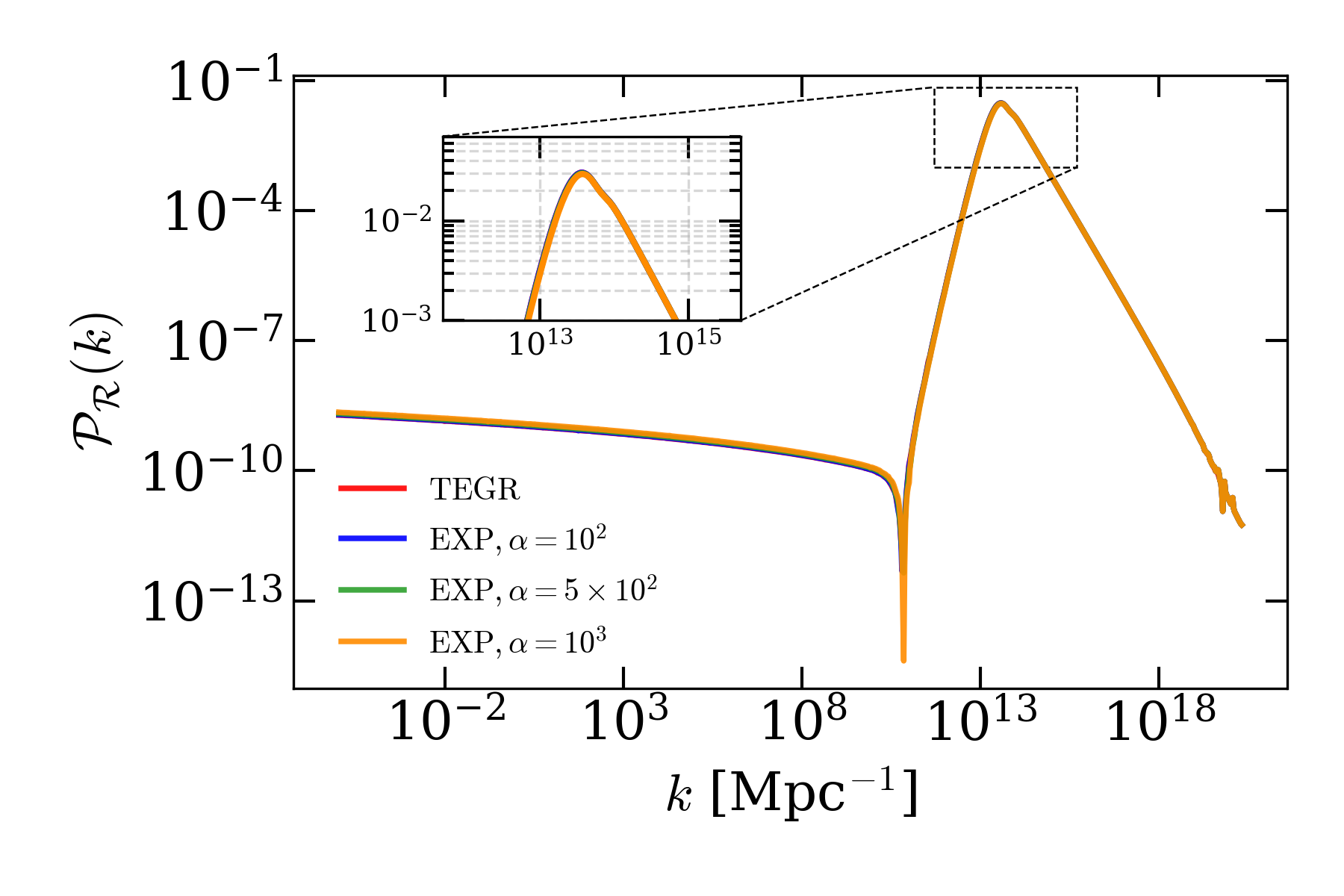}
        
    \end{minipage}

    \caption{Primordial power spectrum for the power-law model across the different fibre inflation configurations. The enhancement at small scales is associated with the ultra slow-roll phase.}
     \label{fig.spectrumPL}
\end{figure}

\begin{figure}[H]
    \centering

    \begin{minipage}[t]{0.48\textwidth}
        \centering
        \hspace{1cm}\textbf{(a) Configuration $V_2$}
        \includegraphics[width=\linewidth]{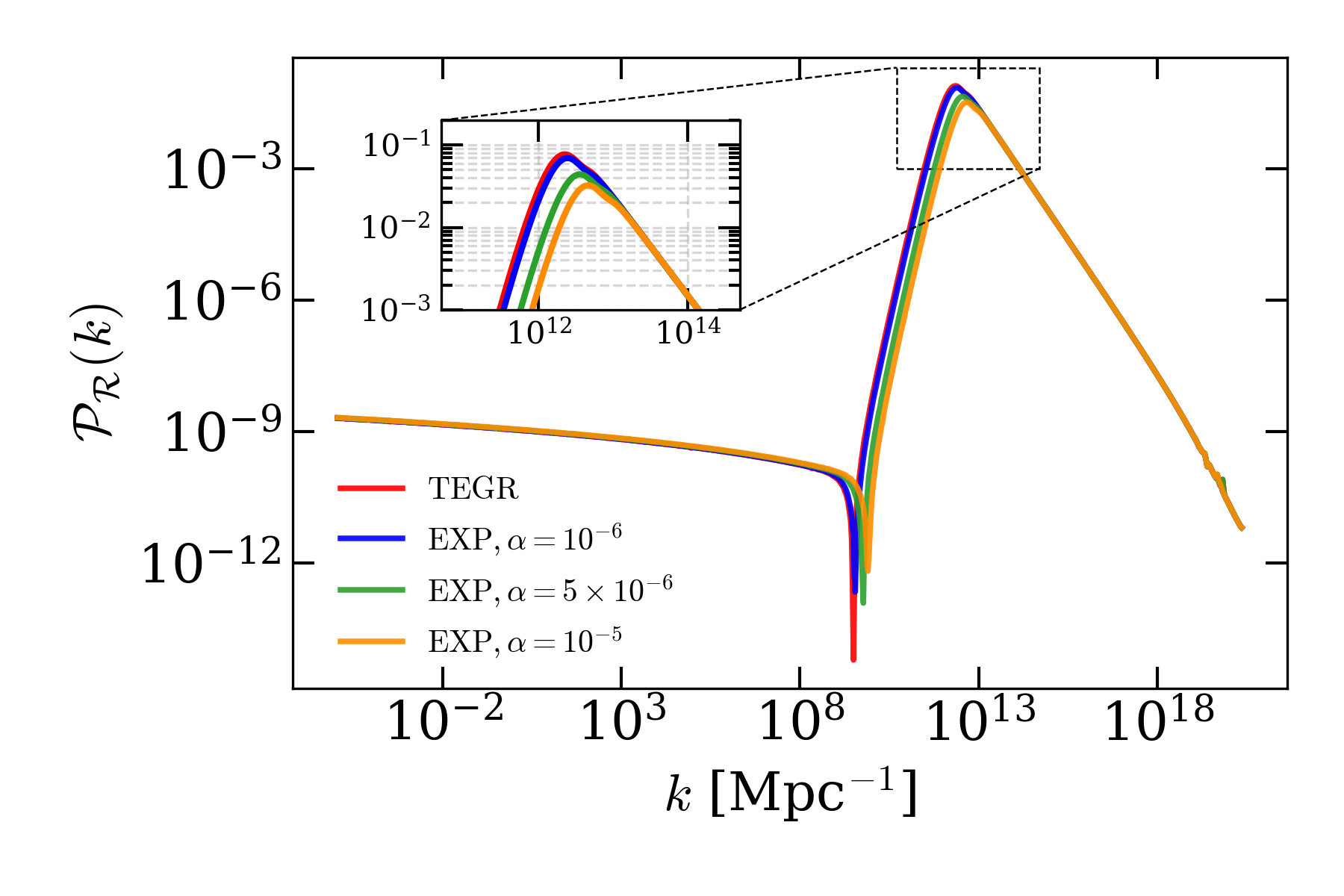}
    \end{minipage}
    \hfill
    \begin{minipage}[t]{0.48\textwidth}
        \centering
        \hspace{1cm}\textbf{(b) Configuration $V_4$}
        \includegraphics[width=\linewidth]{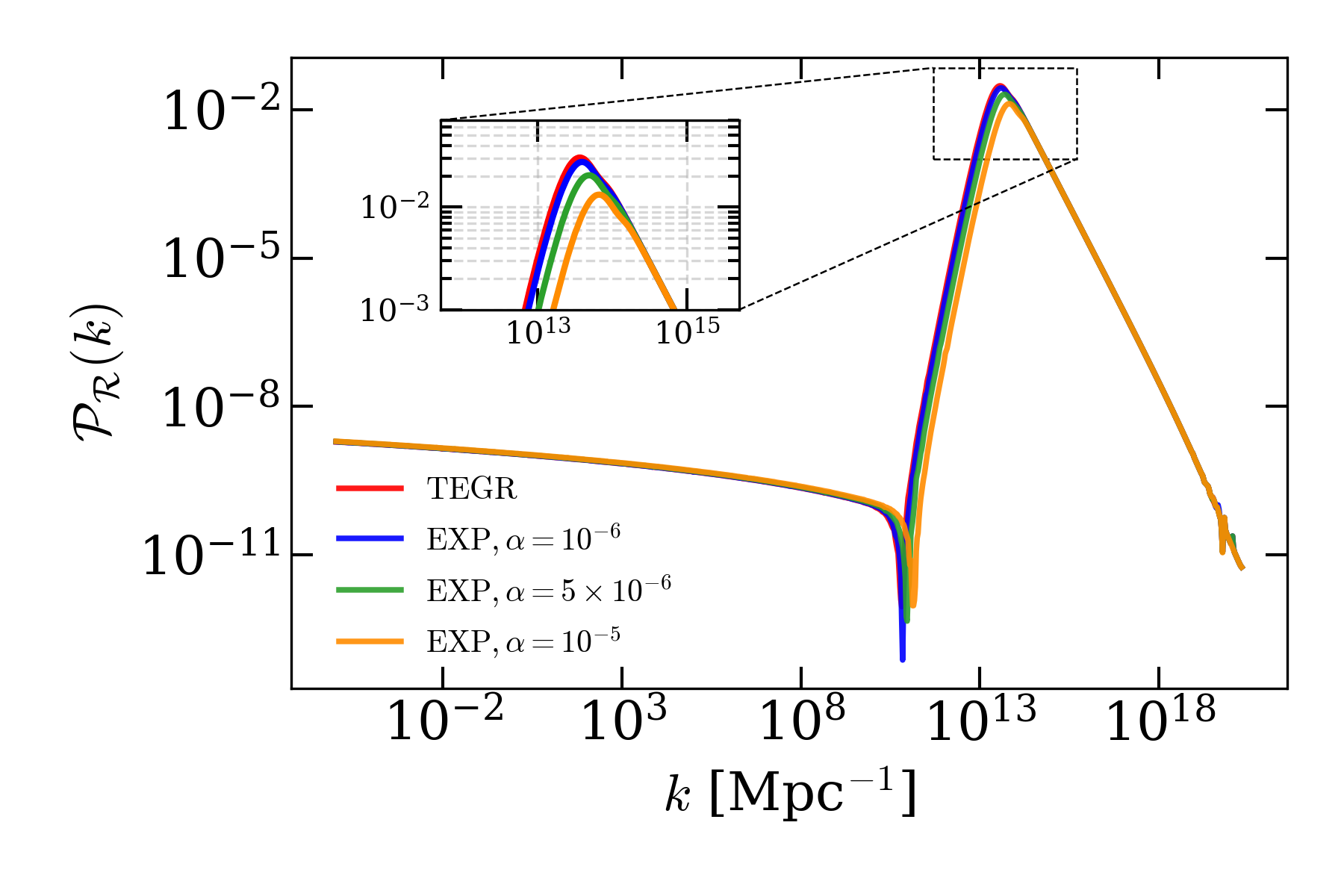}
    \end{minipage}

    \vspace{0.4cm} 

    \caption{Primordial power spectrum for the exponential model for the different fibre inflation configurations, showing the enhancement of scalar perturbations at small scales.}
    \label{fig.spectrumEXP}

\end{figure}

Rewriting Eq. \eqref{mukhanov} in terms of the number of e-folds $N$, using the relation $d/d\tau = \mathcal{H}\,\,d/dN$, the Mukhanov–Sasaki equation in Fourier space takes the form
\begin{equation}
    \frac{d^2v_k}{dN^2} + \left( 1 - \epsilon \right)\frac{dv_k}{dN} + \left[\frac{k^2}{a^2H^2}  + \frac{m^2}{H^2}
     - \left(2-\epsilon+\frac{3}{2}\eta -\frac{1}{2}\epsilon \eta + \frac{1}{4}\eta^2+\frac{1}{2} \frac{d\eta}{dN}\right)\right]v_k = 0 \,.
\end{equation}
This equation governs the evolution of scalar perturbations during inflation and must be solved for each Fourier mode $k$. In the sub-horizon regime, $k \gg aH$, the perturbations behave as free modes in Minkowski spacetime. Accordingly, we impose Bunch–Davies initial conditions \cite{Riotto:2002yw}, given by
\begin{equation}
    v = \frac{1}{\sqrt{2k}} \, ,\quad \frac{dv}{dN} = -i\frac{\sqrt{k}}{\sqrt{2(aH)^2}} \,.
\end{equation}
The primordial power spectrum is then computed from the curvature perturbation $\mathcal{R}_k = v_k/z$ yielding \cite{Sasaki:1986hm,Mukhanov:1988jd,Leach:2000yw}
\begin{equation}
    \mathcal{P}_\mathcal{R}(k) = \frac{k^3}{2\pi^2}\left|\mathcal{R}_k \right|^2 \,.
\end{equation}
In the context of modified $f(T,\phi)$ gravity, the evolution of perturbations is further affected by an effective mass term $m^2$, which arises due to the breaking of local Lorentz invariance. This term introduces additional scale-dependent features in the Mukhanov–Sasaki equation and plays a key role during the USR phase.
For the models considered in this work, the effective mass term takes the following forms. In the Power-law model,
\begin{equation}
m^2= -72 \epsilon H^4 \frac{\alpha \beta(\beta-1)(6H^2)^{\beta-2}}{1+\alpha \beta(6H^2)^{\beta-1}} \,,
\end{equation}
while in the Exponential model it is given by
\begin{equation}
m^2 = - \frac{12\alpha H^4e^{\frac{-H^2}{\beta H_0^2}}\epsilon}{\beta^2H_0^2(1+\frac{\alpha}{\beta}e^{\frac{-H^2}{\beta H_0^2}})} \,.
\end{equation}
The evolution of the effective mass term for both models is shown in Fig.~\ref{fig.massPL} and Fig.~\ref{fig.massEXP}, respectively. As can be seen, the presence of the torsional corrections introduces nontrivial features in $m^2$, which can significantly affect the growth of curvature perturbations during the USR phase.
By numerically solving the Mukhanov–Sasaki equation, we obtain the primordial power spectrum for the different fibre inflation configurations. The resulting spectra for the power-law and exponential models are shown in Fig.~\ref{fig.spectrumPL} and Fig.~\ref{fig.spectrumEXP}, respectively. In all cases, a pronounced enhancement of the power spectrum is observed at small scales, associated with the onset of the USR phase.

Finally, the corresponding peak amplitude $\mathcal{P}_{\mathcal{R}, \text{peak}}$ and the location of the peak $k_{\text{peak}}$ are summarized in Tables~\ref{tabla:powerlaw_spectrum} and \ref{tabla:exponencial_spectrum}. These quantities are of particular importance, as they determine the abundance and mass scale of the produced PBHs.

\begin{table}[H]
\centering
\resizebox{0.7\textwidth}{!}{%
\begin{tabular}{l l  cc cc }
\toprule
\textbf{Model} & \boldmath$\alpha$ 
& \multicolumn{2}{c}{\textbf{Configuration V2}} 
& \multicolumn{2}{c}{\textbf{Configuration V4}} \\
\cmidrule(lr){3-4} \cmidrule(lr){5-6}
& & \boldmath$\mathcal{P}_\mathcal{R\text{,peak}}$ & \boldmath$k_\mathrm{peak}$ 
& \boldmath$\mathcal{P}_\mathcal{R\text{,peak}}$ & \boldmath$k_\mathrm{peak}$ \\
\midrule
TG       & —       & $ 7.71\times10^{-2}$ & $2.16\times10^{12}$ & $3.04\times10^{-2}$ & $3.53\times10^{13}$  \\
PL  & $10^{2}$ & $7.67\times10^{-2} $ & $2.16\times10^{12}$ & $3.02\times10^{-2}$ & $3.53\times10^{13}$ \\

PL  & $5\times10^{2}$ & $7.52\times10^{-2}$ & $2.37\times10^{12}$ & $2.99\times10^{-2} $ & $ 3.87\times10^{13}$ \\
PL  & $10^{3}$ & $7.36\times10^{-2}$ & $2.37\times10^{12}$ & $2.95\times10^{-2}$ & $3.87\times10^{13}$  \\
\bottomrule
\end{tabular}%
}
\caption{Peak amplitude and corresponding scale of the primordial power spectrum for the power-law model, for different values of the parameter $\alpha$ and fibre inflation configurations.}
\label{tabla:powerlaw_spectrum}
\end{table}

\begin{table}[H]
\centering
\resizebox{0.7\textwidth}{!}{%
\begin{tabular}{l l  cc cc }
\toprule
\textbf{Model} & \boldmath$\alpha$ 

& \multicolumn{2}{c}{\textbf{Configuration V2}} 
& \multicolumn{2}{c}{\textbf{Configuration V4}} \\
\cmidrule(lr){3-4} \cmidrule(lr){5-6}
& & \boldmath$\mathcal{P}_\mathcal{R\text{,peak}}$ & \boldmath$k_\mathrm{peak}$ 
& \boldmath$\mathcal{P}_\mathcal{R\text{,peak}}$ & \boldmath$k_\mathrm{peak}$
\\
\midrule
TEGR       & —       & $7.71\times10^{-2} $ & $2.16\times10^{12}$ & $3.04\times10^{-2}$ & $3.53\times10^{13}$ \\
EXP  & $10^{-6}$ & $6.96\times10^{-2}$ & $2.37\times10^{12}$ & $2.76\times10^{-2}$ & $3.87\times 10^{13}$\\
EXP  & $5\times10^{-6}$  & $4.41\times10^{-2}$ & $3.73\times10^{12}$ & $2.04\times 10^{-2}$ & $5.08\times10^{13}$\\
EXP  & $10^{-5}$ & $3.25\times10^{-2}$ & $4.48\times10^{12}$ & $1.33\times10^{-2} $ & $6.67\times 10^{13}$ \\
\bottomrule
\end{tabular}%
}
\caption{Peak amplitude and corresponding scale of the primordial power spectrum for the exponential model, for different values of $\alpha$ and fibre inflation configurations.}
\label{tabla:exponencial_spectrum}
\end{table}

\subsection{Observational Constraints}

The inflationary observables derived from Eqs. \eqref{eq.observables}, evaluated at horizon crossing, receive additional contributions arising from the modified $f(T)$ gravity sector. In particular, the parameters $\eta_R$ and $\delta_{f,T}$ encode deviations from standard single-field inflation, originating from the breaking of local Lorentz invariance and from the slow-roll correction from standard inflation.

These parameters can be expressed as follows:
\begin{itemize}

\item Power-law model

\begin{align}
    \eta_R &= -\frac{24\epsilon\,H^2 \,\alpha\,\beta(\beta-1)\,\left(6H^2\right)^{\beta-2} }{1 + \alpha\,\beta\,\left(6H^2\right)^{\beta-1} }\,,\label{eq: etaR power-law}\\
    \delta_{f_{,T}}&= \frac{  6^{-1+\beta}\,\epsilon\,\alpha\,(\beta-1)\beta\,H^2\big(H^2\big)^{\beta-2}
}{\left[ -1 - \left(6H^2\right)^{-1+\beta}\alpha\,\beta \right] }\,.
\end{align}

\item Exponential model

\begin{align}
    \eta_R &= \frac{
 4\,\alpha\,H\,H'\,
 \exp\,\left(-\frac{H^2}{H_0^2\,\beta}\right)
}{
 H_0^2\,\beta^2\,\left[
 1+\frac{\alpha}{\beta}
 \exp\,\left(-\frac{H^2}{H_0^2\,\beta}\right)
 \right]
 } \label{eq: eta_R EXP}\,,\\
    \delta_{f_{,T}} &= -\,\frac{
 2\,\alpha\,H\,H'\,
 \exp\,\left(-\frac{H^2}{H_0^2\,\beta}\right)
}{
 H_0^2\,\beta^2\,\left[
 1+\frac{\alpha}{\beta}
 \exp\,\left(-\frac{H^2}{H_0^2\,\beta}\right)
 \right]
 } \,.
\end{align}

\end{itemize}

\begin{figure}[H]
    \centering

    \begin{minipage}[t]{0.48\textwidth}
        \centering
        \includegraphics[width=\linewidth]{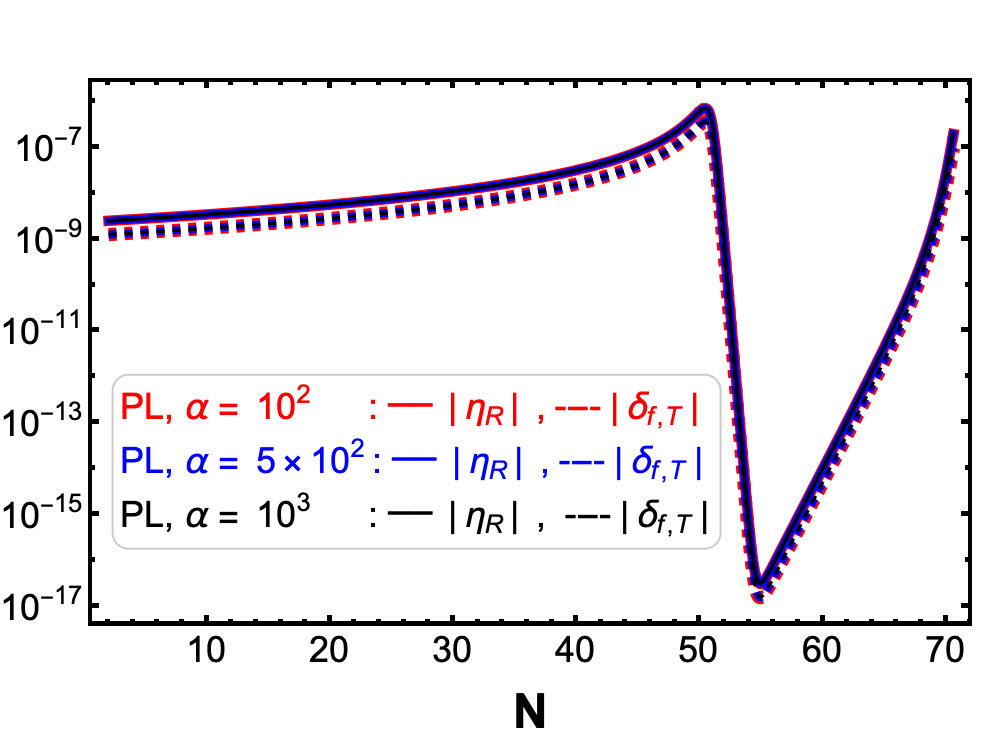}
        \textbf{(a) Configuration $V_2$}        
    \end{minipage}
    \hfill
    \begin{minipage}[t]{0.48\textwidth}
        \centering
        \includegraphics[width=\linewidth]{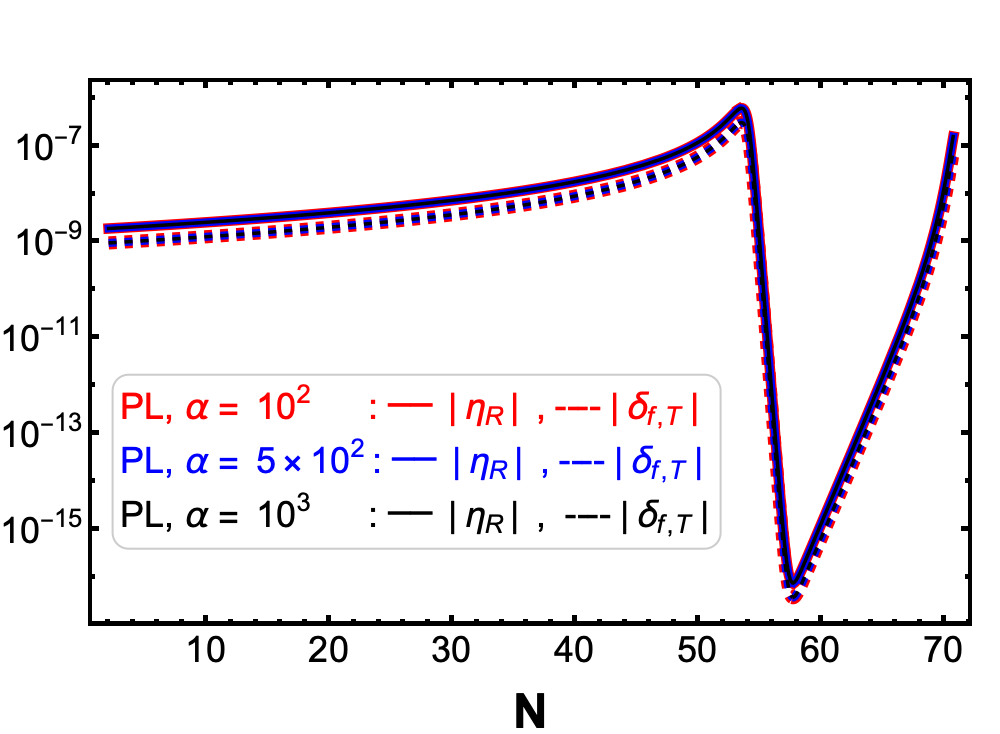}
        \textbf{(b) Configuration $V_4$}        

    \end{minipage}

    \caption{Evolution of the parameters $\eta_R$ and $\delta_{f,T}$ for the Power-law model, for different values of $\alpha$ and fibre inflation potential configurations. These quantities quantify deviations from General Relativity.}
    \label{fig:Parametros-PL}

\end{figure}

\begin{figure}[H]
    \centering

    \begin{minipage}[t]{0.48\textwidth}
        \centering
        \includegraphics[width=\linewidth]{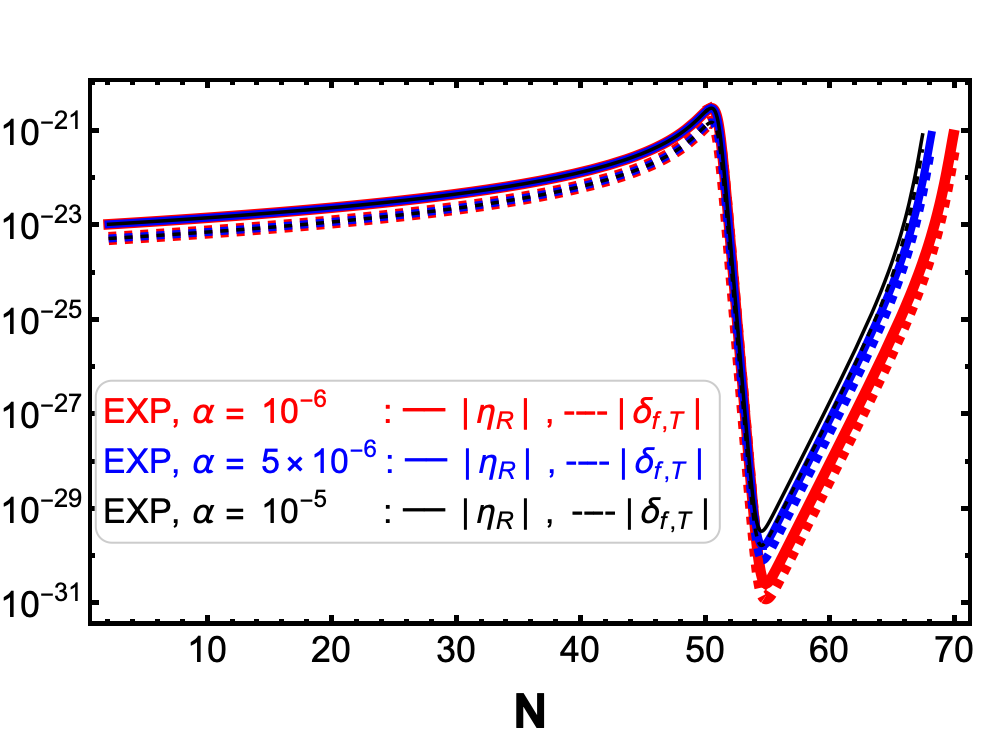}
        \textbf{(a) Configuration $V_2$}        
    \end{minipage}
    \hfill
    \begin{minipage}[t]{0.48\textwidth}
        \centering
        \includegraphics[width=\linewidth]{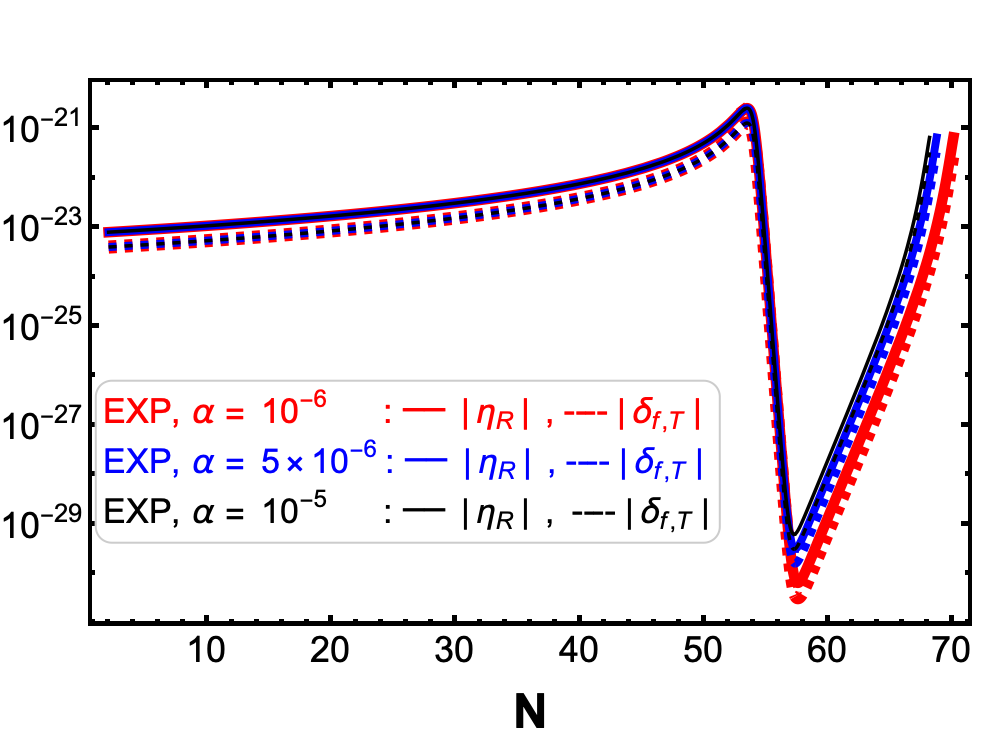}
        \textbf{(b) Configuration $V_4$}        

    \end{minipage}

    \caption{Evolution of the parameters $\eta_R$ and $\delta_{f,T}$ for the Exponential model, for different values of $\alpha$ and fibre inflation potential configurations. These quantities quantify deviations from General Relativity.}
    \label{fig:Parametros-EXP}

\end{figure}

In Figs. \ref{fig:Parametros-PL} and \ref{fig:Parametros-EXP}, we show the evolution of these additional parameters for different values of the correction parameter $\alpha$ and for the various fibre inflation potential configurations. The results are obtained by numerically solving the full background equations without invoking the slow-roll approximation. 

From these figures, it is evident that both $\eta_R$ and $\delta_{f,T}$ remain subdominant throughout the evolution, satisfying $\eta_R, \delta_{f,T} \ll \epsilon, \eta$. This behavior is expected since, in the present setup, we neglect direct couplings between torsion and the scalar field ($f_{,T\phi}=0$), and consider only small deviations from TG.

\begin{figure}[H]
    \centering

    \begin{minipage}[t]{0.48\textwidth}
        \centering
        \includegraphics[width=\linewidth]{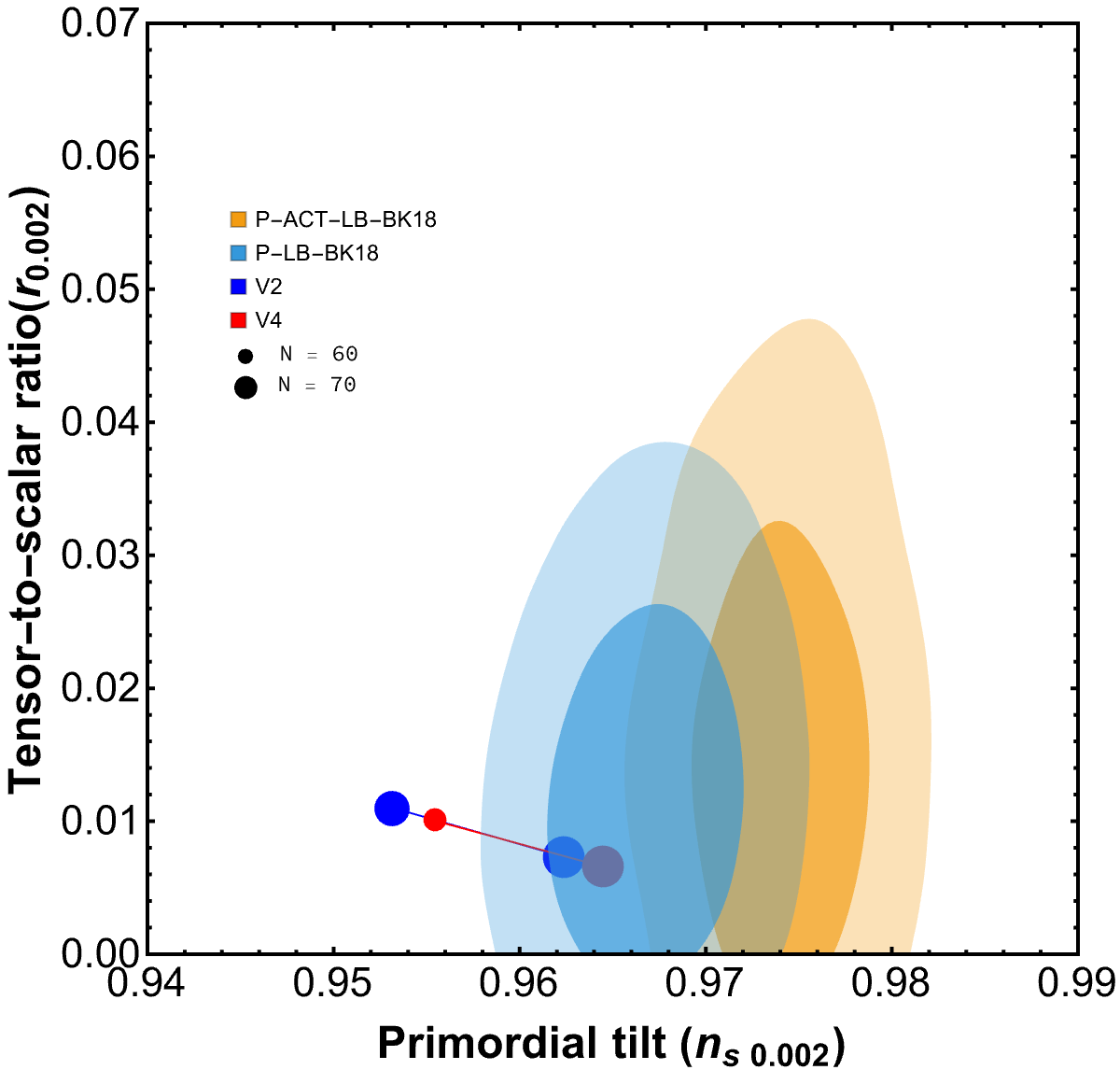}
    \end{minipage}
    \hfill
    \begin{minipage}[t]{0.48\textwidth}
        \centering
        \includegraphics[width=\linewidth]{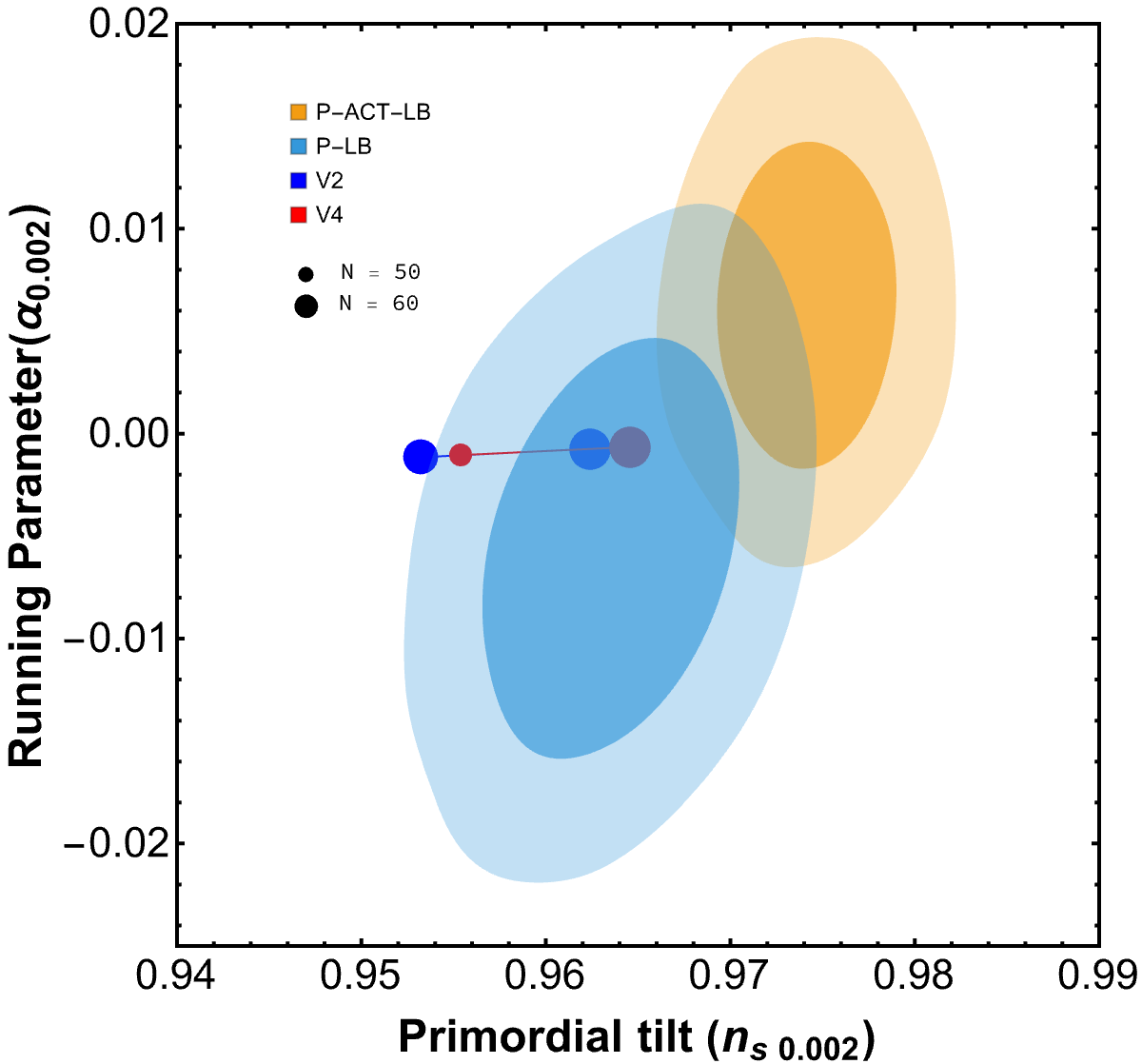}

    \end{minipage}

    \caption{Observational constraints from Planck 2018 on the inflationary observables $(n_s, r)$ \cite{Planck2018}, ACT-2025 \cite{AtacamaCosmologyTelescope:2025blo} and BICEP/Keck 2021 \cite{BICEP:2021xfz}. The small, medium and large circles correspond to N = 60 and 70 e-folds, respectively. The light-shaded and dark-shaded regions correspond to the 95 $\%$CL and 68 $\%$CL, respectively.}

    \label{fig: OBScontraint}

\end{figure}

As a consequence, the standard expressions for inflationary observables are approximately recovered, with small corrections from the torsional sector:
\begin{align}
   n_s &=-2\epsilon-\eta_Q+2\eta_R \,,\\
    r &= 16(\epsilon-\delta_{f,T})\,,\\
    \alpha_s &= \frac{d n_s}{d\ln k} = 2 \epsilon \eta_Q-\eta_Q' -\eta_{R}'\,. \label{eq.observables}
\end{align}
In Fig. \ref{fig: OBScontraint}, we present the observational constraints in the $(n_s, r)$ plane for the  fibre inflation potential configurations, evaluated within the slow-roll approximation in the TG limit. The predictions corresponding to $V_2$, and $V_4$ lie within the $68\%$ and $95\%$ confidence level regions of the Planck 2018 data \cite{Planck2018}, compared to the ACT observational data \cite{AtacamaCosmologyTelescope:2025blo} and the constraints in the $n_s -\alpha_s$ plane with current CMB observations. This potential is in tension with the confidence regions. Complementarily, in Table \ref{tab:nsr_modelos} we report the numerical values of the inflationary observables $(n_s, r, \alpha_s)$ for different numbers of e-folds ($N=60, 70$).

\begin{table}[H]
\centering
\begin{tabular}{c c c}
\toprule
\textbf{Model} & \multicolumn{2}{c}{\textbf{Observables $(n_s, r,\alpha_s)$}} \\
\cmidrule(lr){2-3} 
 & \textbf{N=60} & \textbf{N=70} \\
\midrule
$V_2$ & $(0.953, 0.011, -0.0011)$ & $(0.962, 0.007, -0.0007)$  \\
$V_4$ & $(0.955, 0.010, -0.0010)$ & $(0.964, 0.006, -0.0006)$ \\
\bottomrule
\end{tabular}
\caption{Inflationary observables $(n_s, r, \alpha_s)$ for the different fibre inflation potential configurations, evaluated at $N=60$, and $70$ e-folds.}
\label{tab:nsr_modelos}
\end{table}

\subsection{Primordial Black Hole Mass}

The mass of PBHs formed during the radiation-dominated era is determined by the mass enclosed within the horizon at the time of collapse. In this context, the PBH mass can be expressed as a fraction of the horizon mass 
\begin{equation}
    M_{\text{PBH}} = \gamma M_\text{H} = \gamma \frac{4}{3}\pi \rho H^{-3} \,,
\end{equation}
where $\gamma$ parametrizes the efficiency of gravitational collapse and is typically taken as $\gamma \simeq 0.2$ during radiation domination \cite{Ballesteros:2017fsr}.
Since PBHs are produced when enhanced curvature perturbations re-enter the horizon, their mass is directly related to the scale of the perturbation. In the radiation era, the energy density scales as $\rho_r \propto g(T) T^4$, allowing one to relate the PBH mass to the comoving wavenumber $k$ as \cite{Ballesteros:2020qam,Ozsoy:2018flq}
\begin{equation}
    M(k) = \gamma M_H \left( \frac{ g(T_f)}{g(T_{eq})}\right)^{1/3}\left( \frac{ g_s(T_f)}{g_s(T_{eq})}\right)^{-2/3}\left( \frac{k}{k_{eq}}\right)^{-2} \,.
\end{equation}
Assuming that the relativistic degrees of freedom for the energy and entropy densities coincide, i.e. $g(T) = g_s(T)$, we take $g(T_{eq}) = 3.38$ at matter-radiation equality and $g(T) = 106.75$ for the Standard Model at high temperatures. The corresponding horizon mass at equality is $M_H \simeq 5.6\times 10^{50}\text{g}$.
In this regime, PBH formation occurs well after inflation, during standard radiation domination, so modifications to gravity can be safely neglected and the standard relation between horizon mass and scale applies. Physically, this reflects the fact that PBHs form from the collapse of horizon-sized overdensities, making their mass proportional to the horizon mass at re-entry.

For practical purposes, the relation between PBH mass and the scale of the curvature perturbations can be approximated as \cite{Ballesteros:2020qam} 
\begin{equation}
    M(k) \simeq 10^{17}\left(\frac{k}{2\times 10^{14 }\text{Mpc}^{-1}} \right) ^{-2} \text{g} \,,
\end{equation}
which provides a direct link between features in the primordial power spectrum and the resulting PBH mass spectrum.

\subsection{Primordial black holes in radiation epoch }
The production of PBHs can be estimated using the Press–Schechter formalism, where sufficiently large density perturbations collapse upon horizon re-entry. This process depends crucially on the threshold value $\delta_c$, which characterizes the minimum density contrast required for gravitational collapse.

Assuming Gaussian statistics for the primordial curvature perturbations, the fraction of PBHs contributing to dark matter can be written as
\begin{equation}
    f_{\text{PBH}} = \frac{\Omega_{\text{PBH}}}{\Omega_{\text{DM}}}\simeq \left(\frac{M}{10^{18} \text{g}} \frac{1.25\times 10^{15}}{\sqrt{2\pi \sigma^2(M)}} \int_{\delta_c}^{\infty} \exp \left[ -\frac{\delta^2}{2\sigma^2(M)} \right] d\delta\right) \,,
\end{equation}
which depends exponentially on the ratio between the threshold $\delta_c$ and the variance of the density perturbations $\sigma^2(M)$. The latter is obtained by integrating the primordial power spectrum
\begin{equation}
    \sigma^2(M)=\frac{16}{81}\int_0^{\infty}\frac{dq}{q}\left( \frac{q}{k}\right)^4\mathcal{P}_\mathcal{R}(q)W^2\left(\frac{q}{k}\right) \,.
\end{equation}
In this work, we adopt a Gaussian window function $W(x)=\exp(-x^2/2)$, which smooths the perturbations on the relevant scale. It is worth emphasizing that the resulting PBH abundance is highly sensitive to the choice of window function and to the amplitude of the power spectrum, leading to potentially large variations in $f_{\text{PBH}}$.

\begin{figure}[H]
    \centering

    \begin{minipage}[t]{0.48\textwidth}
        \centering
        \hspace{1cm}\textbf{(a) Configuration $V_2$}
        \includegraphics[width=\linewidth]{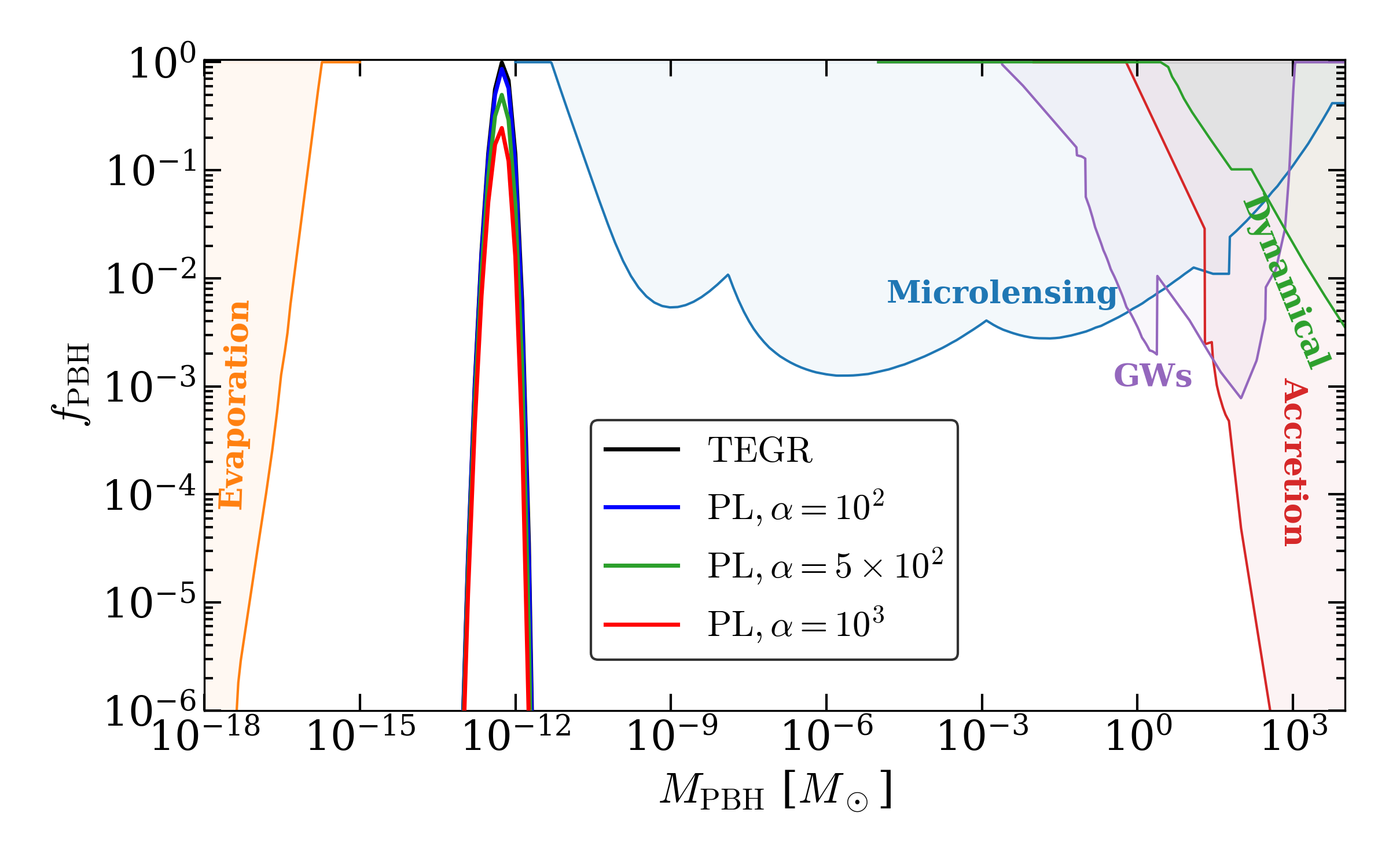}
    \end{minipage}
    \hfill
    \begin{minipage}[t]{0.48\textwidth}
        \centering
        \hspace{1cm}\textbf{(b) Configuration $V_4$}
        \includegraphics[width=\linewidth]{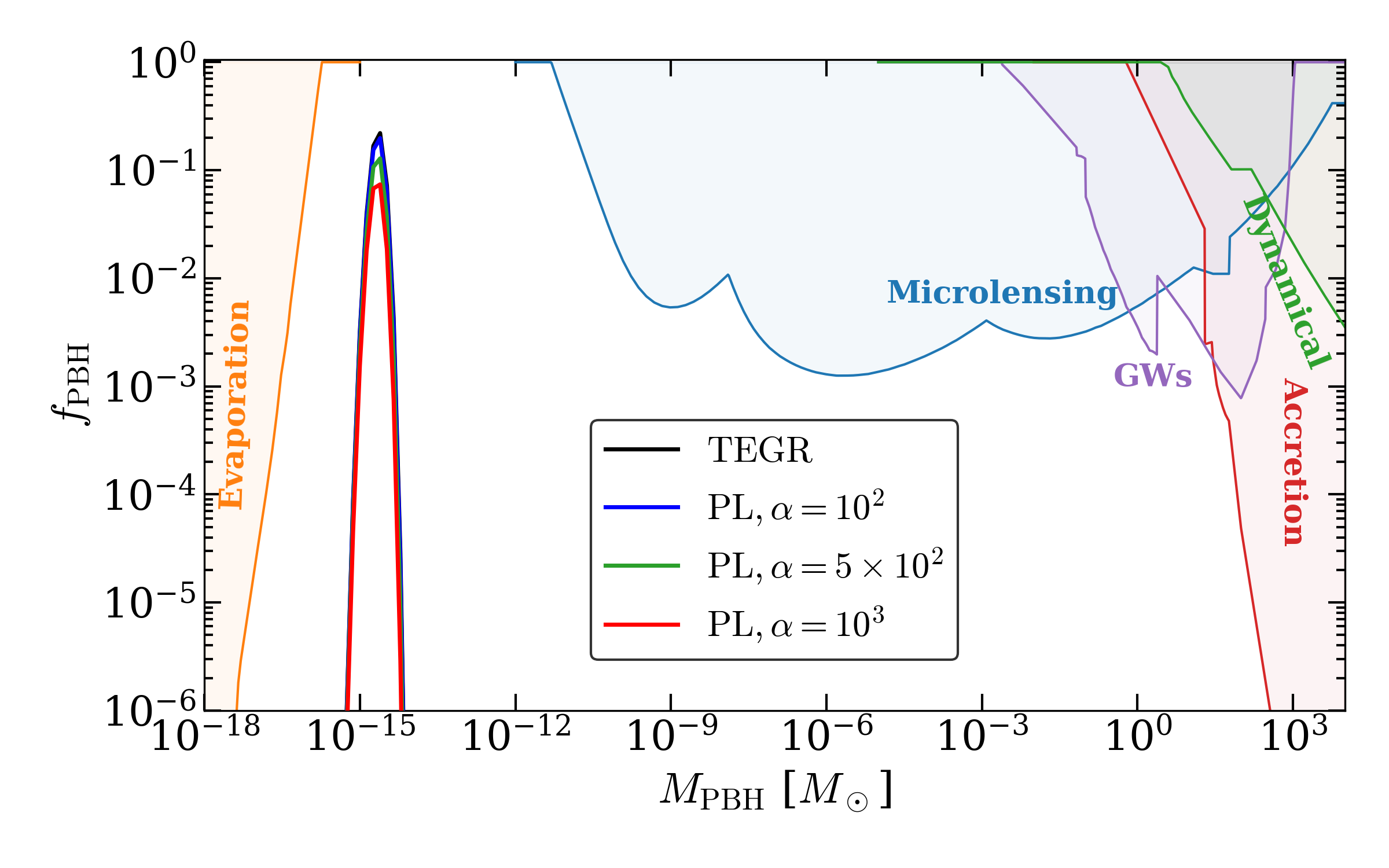}
    \end{minipage}

    \caption{Fraction of PBHs contributing to dark matter for the Power-Law model, computed for the different fibre inflation potential configurations $V_2$ and $V_4$.}
    \label{fig:FPBHS-PL}

\end{figure}

\begin{figure}[H]
    \centering

    \begin{minipage}[t]{0.48\textwidth}
        \centering
        \hspace{1cm}\textbf{(a) Configuration $V_2$}
        \includegraphics[width=\linewidth]{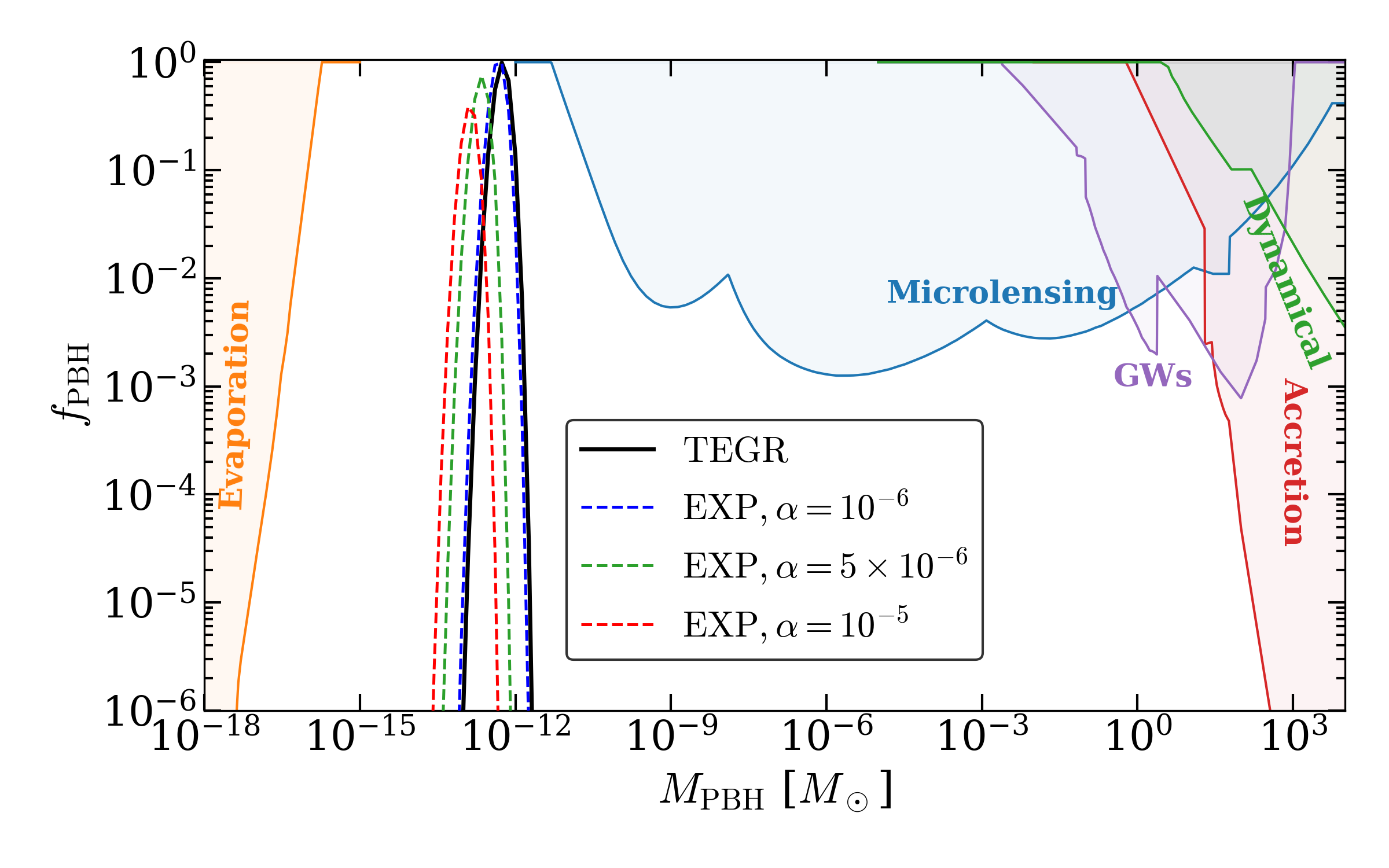}
    \end{minipage}
    \hfill
    \begin{minipage}[t]{0.48\textwidth}
        \centering
        \hspace{1cm}\textbf{(b) Configuration $V_4$}
        \includegraphics[width=\linewidth]{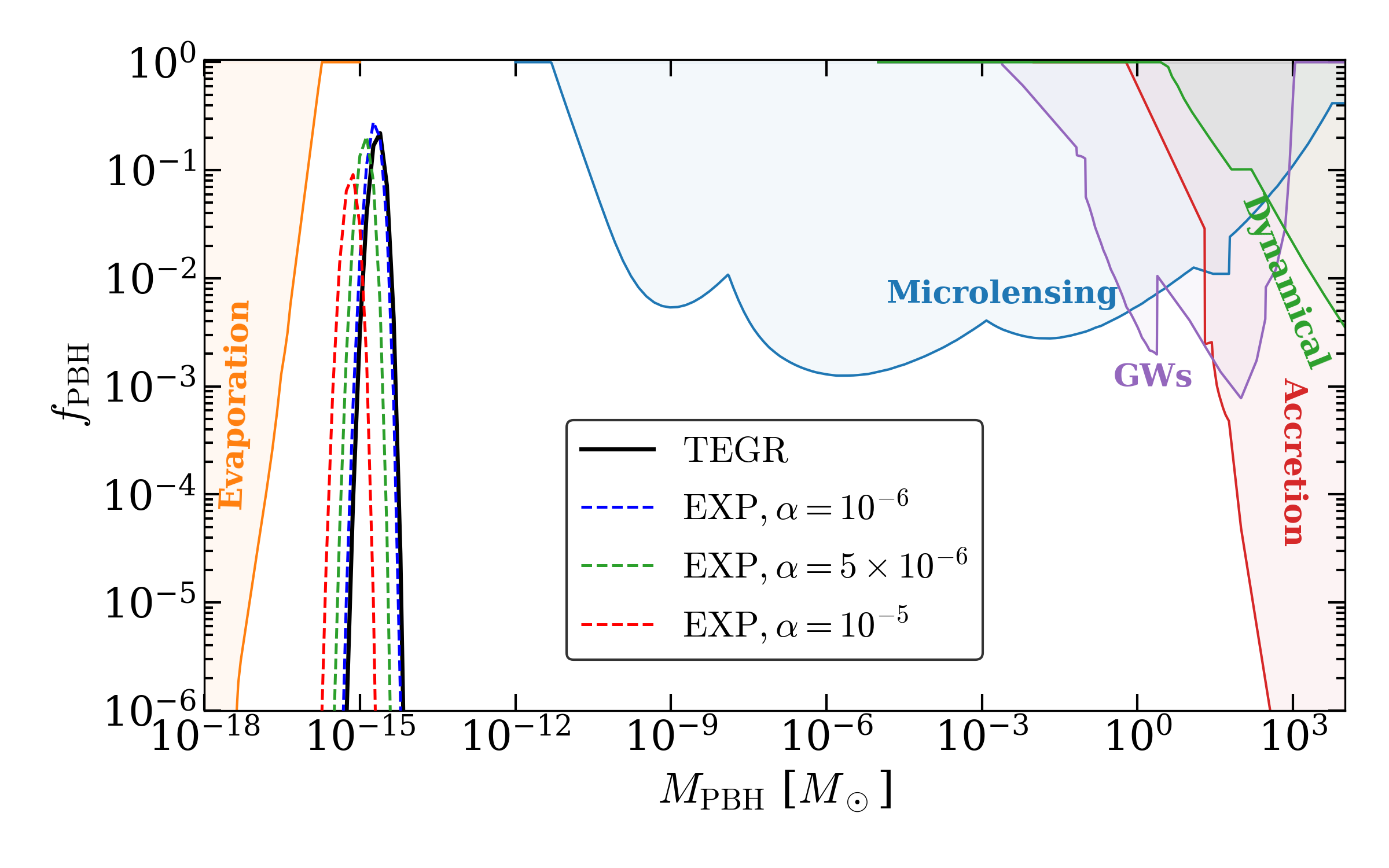}
    \end{minipage}

    \caption{Fraction of PBHs contributing to dark matter for the Exponential model, computed for the different fibre inflation potential configurations $V_2$ and $V_4$.}
    \label{fig:FPBHS-EXP}

\end{figure}

In Figs. \ref{fig:FPBHS-PL} and \ref{fig:FPBHS-EXP} we show the resulting PBH dark matter fraction for the different configurations of the fibre inflation potential in both the Power-Law and Exponential models. The corresponding numerical values, including the peak amplitude of the power spectrum, the PBH mass scale, and the threshold parameter $\delta_c$, are summarized in Tables \ref{tab.V2V4PBHPOWERLAW} and \ref{tab.V2V4PBHEXP}.

\begin{table}[H]
\centering
\scriptsize
\begin{tabular}{c | c c c c | c c c c}
\toprule
\textbf{\raisebox{-1.5ex}{Model}} & 
\multicolumn{4}{c|}{\textbf{Configuration V$_2$}} & 
\multicolumn{4}{c}{\textbf{Configuration V$_4$}} \\
\cline{2-9}  
 & $\mathcal{P_R}$ & $M_\text{max}/M_\odot$ & $\delta_c$ & $f_\text{PBH}$ &
   $\mathcal{P_R}$ & $M_\text{max}/M_\odot$ & $\delta_c$ & $f_\text{PBH}$ \\
\midrule
TEGR & $7.71\times10^{-2}$ & $5.47\times10^{-13}$ & $0.60$ & $0.99$ & $3.04\times10^{-2}$ & $2.46\times10^{-15}$ & $0.40$ & $0.22$ \\
PL, $\alpha=10^{2}$ & $7.67\times10^{-2}$ & $5.47\times10^{-13}$ & $0.60$ & $0.86$ & $3.02\times10^{-2}$ & $2.46\times10^{-15}$ & $0.40$ & $0.19$ \\
PL, $\alpha=5\times10^{2}$ & $7.52\times10^{-2}$ & $5.47\times10^{-13}$ & $0.60$ & $0.49$ & $2.99\times10^{-2}$ & $2.46\times10^{-15} $ & $0.40$ & $0.12$ \\
PL, $\alpha=10^{3}$ & $7.36\times10^{-2}$ & $5.47\times10^{-13}$ & $0.60$ & $0.24$ & $2.95\times10^{-2}$ & $2.46\times10^{-15}$ & $0.40$ & $0.07$ \\
\bottomrule
\end{tabular}
\caption{Results of the fractions of PBHs as dark matter 
for the Power-Law model using configurations $V_2$ and $V_4$.}
\label{tab.V2V4PBHPOWERLAW}
\end{table}

\begin{table}[H]
\centering
\scriptsize
\begin{tabular}{c | c c c c | c c c c}
\toprule
\textbf{\raisebox{-1.5ex}{Model}} & 
\multicolumn{4}{c|}{\textbf{Configuration V$_2$}} & 
\multicolumn{4}{c}{\textbf{Configuration V$_4$}} \\
\cline{2-9}  
 & $\mathcal{P_R}$ & $M_\text{max}/M_\odot$ & $\delta_c$ & $f_\text{PBH}$ &
   $\mathcal{P_R}$ & $M_\text{max}/M_\odot$ & $\delta_c$ & $f_\text{PBH}$ \\
\midrule
TEGR & $7.71\times10^{-2}$ & $5.47\times10^{-13}$ & $0.60$ & $0.99$ & $3.04\times10^{-2}$ &  $2.46\times10^{-15}$ & $0.40$ &  $0.22$ \\
EXP, $\alpha=10^{-6}$ & $ 6.96\times 10^{-2} $ &  $5.47\times10^{-13}$ & $0.57$&  $0.98$ & $2.76\times10^{-2}$ &  $1.82\times10^{-15} $ & $0.38$ &  $0.28$ \\

EXP, $\alpha=5\times10^{-6}$ &  $4.41\times10^{-2}$ &  $2.22\times10^{-13}$ & $0.46$ & $0.74$ &
                                 $2.04\times10^{-2} $ &  $1.35\times10^{-15}$ & $0.33$ &  $0.20$ \\

EXP, $\alpha=10^{-5}$ &  $3.25\times10^{-2}$ &  $1.22\times10^{-13}$ & $0.40$ &  $0.38$ &
                          $1.33\times10^{-2}$ &  $7.39\times10^{-16}$ & $0.27$ &  $0.09$ \\
\bottomrule
\end{tabular}
\caption{Fraction of PBHs as dark matter of the Exponential model, considering the configurations $V_2$ and $V_4$.}
\label{tab.V2V4PBHEXP}
\end{table}
\vspace{-0.5cm}

\section{Conclusions}\label{conclusions}

In this work, we have studied inflationary dynamics and primordial black hole (PBH) production within the framework of modified teleparallel gravity, focusing on scalar–torsion models of the form $f(T,\phi)=f(T)-V(\phi)$. We considered two representative modifications of teleparallel gravity, namely the power-law and exponential models, together with a fibre inflation potential capable of generating a transient ultra slow-roll (USR) phase.

A key ingredient for PBH formation is the enhancement of the primordial curvature power spectrum during the USR phase. In order to accurately capture this effect, we solved the full Mukhanov–Sasaki equation without relying on the slow-roll approximation. This is essential since PBH production is highly sensitive to departures from slow-roll dynamics.

We found that modified teleparallel gravity introduces non-trivial corrections both at the background and perturbation levels. In particular the first slow-roll parameter $\epsilon$ is modified by torsional contributions, which can either enhance or suppress its evolution depending on the model parameters. The Mukhanov–Sasaki equation acquires an effective mass term $m^2$, originating from the breaking of local Lorentz invariance, which directly affects the growth of curvature perturbations. Additional slow-roll parameters, such as $\eta_R$ and $\delta_{f,T}$, appear as corrections to the standard inflationary observables. However, in the absence of scalar–torsion couplings, these contributions remain subdominant, allowing a recovery of the standard predictions for $n_s$, $r$, and $\alpha_s$.

From the numerical analysis, we showed that the fibre inflation potential successfully generates a transient USR phase, leading to a significant amplification of the power spectrum. The resulting peak amplitude and its corresponding scale depend sensitively on both the potential parameters and the modification parameter $\alpha$. This enhancement translates into PBH formation when the corresponding modes re-enter the horizon during the radiation-dominated era. In this regime, the PBH mass is related to the horizon mass and scales with the comoving wavenumber as $M(k) \propto k^{-2}$, which is a characteristic feature of radiation domination. Consequently, the location of the peak in the power spectrum directly determines the PBH mass scale.

Our results show that in the power-law model, increasing $\alpha$ suppresses the depth of the USR phase, reducing the enhancement of the power spectrum and therefore decreasing the PBH abundance. In contrast, the exponential model exhibits a richer phenomenology, where different values of $\alpha$ can either suppress or enhance PBH production, depending on the interplay between the exponential corrections and the inflationary dynamics.

The computed PBH abundance, obtained using the Press–Schechter formalism under the assumption of Gaussian perturbations, indicates that certain configurations of the fibre inflation potential can produce a significant fraction of PBHs contributing to dark matter. The abundance is extremely sensitive to the amplitude of the power spectrum and the threshold parameter $\delta_c$, highlighting the importance of precise modeling of the inflationary phase.
Modified teleparallel corrections can shift both the peak amplitude and the corresponding PBH mass, thereby affecting the viability of PBHs as dark matter candidates.

Overall, our analysis demonstrates that modified teleparallel gravity provides a viable and rich framework for PBH production, with distinctive signatures arising from torsional effects. While the corrections to standard inflationary observables remain small, their impact on the dynamics of perturbations and PBH formation can be significant.

Future work could explore the role of nonminimal scalar–torsion couplings, non-Gaussianities, and reheating effects, which may further enhance or constrain PBH production in these models.

\begin{acknowledgments}
Y. V. acknowledges the financial support of DIDULS/ULS, through the project Nº PTE2318.

\end{acknowledgments}

\bibliographystyle{spphys}   
\bibliography{bio}

\end{document}